\documentclass[12pt]{iopart}

\usepackage{iopams}
\usepackage{graphicx}
\graphicspath{
{./}
{/home/jripley/SphericalCollapse/EdGB_Exp/plt_files_for_paper/longer_paper/}
}
\usepackage{color}
\usepackage{subfig}

\begin{document}

\title[Gravitational Collapse in Einstein Dilaton-Gauss-Bonnet Gravity]
{Gravitational Collapse in Einstein Dilaton Gauss-Bonnet Gravity}

\author{Justin L Ripley and Frans Pretorius}

\address{
 Department of Physics, Princeton University, Princeton, New Jersey 08544, USA.
}
\ead{jripley@princeton.edu and fpretori@princeton.edu}
\vspace{10pt}
\begin{indented}
\item[]February 2019
\end{indented}

\begin{abstract}
	We present results from a numerical study of spherical gravitational
collapse in 
shift symmetric Einstein dilaton Gauss Bonnet (EdGB)
gravity. This modified gravity theory has a single coupling parameter
that when zero reduces to general relativity (GR) minimally coupled
to a massless scalar
field. We first show results from the weak EdGB coupling limit,
where we obtain solutions that smoothly
approach those of the Einstein-Klein-Gordon system of GR. Here,
in the strong field
case, though our code does not utilize horizon penetrating coordinates,
we nevertheless find tentative evidence that approaching black hole formation
the EdGB modifications cause the growth of scalar field ``hair'', consistent
with known static black hole solutions in EdGB gravity.
For the strong EdGB coupling regime, in a companion paper we
first showed results that even in the weak field (i.e. far from
black hole formation), the EdGB equations are of mixed type:
evolution of the initially hyperbolic system of partial differential
equations lead to formation of a region where their character
changes to elliptic. Here, we present more details about this regime.
In particular, we show that an effective energy density based on 
the Misner-Sharp mass is negative near these elliptic regions, 
and similarly the null convergence condition is violated then.
\end{abstract}

%
\vspace{2pc}


\section{\label{sec:introduction}Introduction}	

While General Relativity (GR) has passed all experimental and
observational tests so far (caveats with dark energy and dark matter
aside), there are well known reasons to suspect that GR is not a complete
theory of gravity. One reason is that at the level of the classical
equations of motion,
black hole (BH) and most cosmological solutions are geodesically 
incomplete \cite{hawking1975large}, with the expectation that
these spacetimes also generically contain curvature singularities.
Another is that as matter is quantum in nature, the Einstein
field equations relating a classical description of geometry
to a classical stress energy tensor of matter can 
only be an approximate theory; though the
predominant opinion today seems to be that the
resolution of this issue is that geometry is also fundamentally
quantum 
(as opposed to, for example, a completely novel theoretical construct that
reduces to the modern, tested physical theories in appropriate limits),
there is no evidence for this at present. Though
regardless of theoretical reasons to think GR is incomplete,
from a purely empirical point of view 
we have only recently entered the era where we can begin to verify 
the dynamical strong field predictions of GR, through gravitational
wave (GW) observation of compact object mergers. Though initial
tests are consistent with the GR description of these
events~\cite{LIGOScientific:2019fpa},
we are still in the early days of GW astronomy, and the data
cannot yet provide high precision tests of this regime.

One problem with achieving the tightest possible constraints
on deviations from GR in the strong field (or discovering them),
is at present we have no {\em interesting, viable} alternatives to GR
that can give quantitative predictions to the analogue of the merger regime
of BH inspiral in GR. This is where the predominant
share of SNR (signal-to-noise ratio) is coming from with current
detections (in particular GW150914), and where one might expect
to see the first hints of corrections to GR. By {\em interesting}
we mean theories that when restricted to regimes that
are consistent with existing non-GW tests nevertheless
still offer significant differences for BH mergers; by {\em viable}
we mean theories that possess a well-posed initial value
problem (IVP) that can be solved to make predictions
of mergers to confront with data. For example, a class of
viable but uninteresting
theories in this regard (and we emphasize we certainly do not mean
``uninteresting'' for any other reason) are the typical scalar tensor
theories, such as Brans-Dicke, as they have the same vacuum sector
of solutions, hence BH mergers, as GR. Another example
that is viable but {\em likely} uninteresting is
Einstein-Maxwell-Dilaton gravity; 
here the problem is to obtain mergers with noticeable deviations in the
GW emission requires what is expected to be astrophysically unrealistic
amounts of electric charge for the BHs~\cite{Hirschmann:2017psw}.

	Over the past several years
two likely {\em interesting} modified gravity theories have attracted
the attention of researchers attempting to study the full non-linear
BH merger problem~\cite{Okounkova:2017yby,Okounkova:2018pql,
Benkel:2016kcq,Witek:2018dmd}: dynamical Chern-Simons (dCS) gravity
(see e.g. \cite{Alexander:2009tp}),
and Einstein-dilation-Gauss-Bonnet (EdGB)
gravity(see e.g. \cite{Kanti:1995vq,Maeda:2009uy,Yagi:2015oca}),
the latter being
the focus of this paper. One the main motivations
of these studies, and ours here, is to understand
how in principle strong field merger dynamics could differ from
GR (as opposed to any observational or theoretical impetus arguing
for such modifications on the scale of astrophysical BHs).
In terms of testing GR though BH mergers, 
what is interesting
about the particular variant of EdGB gravity we consider here is
it does not admit the Schwarzschild or Kerr BH solutions
of GR. Instead, the analogue BH solutions only exist above
a minimum length scale related to the coupling constant $\lambda$
in the theory, and feature scalar
``hair''~\cite{Kanti:1995vq,Sotiriou:2013qea,Sotiriou:2014pfa}.
Moreover, for values of $\lambda$ that would produce significant changes
in stellar mass BHs, the corresponding effect on material compact objects
such as neutron stars is insignificant~\cite{Yagi:2015oca}, implying
this theory
could be consistent with current GR tests, yet give different results for
stellar mass BH mergers.

One problem
with EdGB gravity relates to whether it is viable in the
above sense of the word. This paper is a follow up to a
first study~\cite{PhysRevD.99.084014} of gravitational collapse in EdGB
gravity in spherical
symmetry to begin to address this issue from the level of
fully non-linear dynamical solutions. Earlier work on the well-posedness
of EdGB gravity~\cite{Papallo:2017qvl,Papallo:2017ddx} 
considered the linearized equations in the small coupling parameter limit,
and found that these equations are at best weakly hyperbolic
about generic backgrounds, within a class of ``generalized harmonic'' gauges.
This is certainly a sign for concern, however demanding
that a theory be well-posed in all possible situations might
be unnecessarily restrictive if problems do not occur in scenarios
of interest, here in particular for binary BH mergers. Considering
the recent results of \cite{Kovacs:2019jqj}, we also mention
there may exist other gauges for which EdGB may have well posed
initial value problem for generic small field initial data. 

Given how challenging
solving for BH merger spacetimes is in GR alone, it makes
sense to tackle this problem in EdGB gravity beginning with simpler scenarios
that capture some aspects of the final problem, uncover
any issues that might arise, and if there are no show-stoppers,
move forward. One approach along these lines follows an 
effective field theory interpretation of EdGB gravity, beginning
with the GR solution, then examining perturbative corrections.
The advantage to this approach is one can begin with fully non-linear
GR BH merger solutions.
The first step here is the so-called decoupling limit
(see \ref{sec:static_decoupled}),
where the EdGB scalar is not yet allowed to back-react on the geometry;
this has successfully been carried out in
~\cite{Witek:2018dmd} (a similar approach has been taken in dCS
gravity in~\cite{Okounkova:2017yby}, and even recently
extended to first order metric perturbations~\cite{Okounkova:2018pql}).

Another approach, that we follow here, is to begin with the
fully non-linear, non-perturbative EdGB equations, but in a
symmetry reduced setting. The benefit of this is we can
immediately begin looking for non-perturbative deviations
from the predictions of GR. The natural, simplest symmetry to consider 
for our purposes is spherically
symmetry, as this allows us to study black hole formation in asymptotically
flat, 4-dimensional spacetimes. One key result of our initial
study, described in~\cite{PhysRevD.99.084014}, is within this
symmetry class we do identify a regime of EdGB gravity that is ``pathological''
from the perspective of having a well-posed IVP :
specifically, in the strong coupling regime, we find scenarios
where evolution of initial data leads to the EdGB dilaton equation
changing character from hyperbolic to elliptic within a region of
the spacetime (or said another way, this equation is
then actually of mixed type).
However, as discussed more in~\cite{PhysRevD.99.084014}, given how this
phenomenon scales with the magnitude of the coupling parameter,
there are regimes of EdGB gravity that may yet offer
a viable, interesting modified gravity scenario for application
to GW astronomy (the main limitation of this first study,
aside from symmetry considerations, is since we do not use horizon
penetrating coordinates, we cannot address the long time, non-linear
stability of BH's regardless of the magnitude of the EdGB coupling parameter).

We should also mention that variants of EdGB gravity have been
extensively studied in the cosmological context. There is a vast literature
on this, and since this topic is outside of our scope,
we will not attempt to cite the relevant papers,
but instead refer the reader to the recent
review articles~\cite{Quiros:2019ktw,Kobayashi:2019hrl} for further reference.
In cosmology the equations are usually analyzed in the form of Horndeski
theories (roughly Einstein gravity coupled to a scalar field
with all possible non-standard kinetic terms that yield second
order equations of motion), with EdGB gravity being a sub-class of the
most general Horndeski theory. The motivations there are
more often driven by the need to explain dark energy, or to come
up with models of the early universe, including inflation or
bouncing models. What is particularly interesting with regard
to the last-named problem is some Horndeski theories can violate
the null convergence condition (NCC, or equivalently the null
energy condition (NEC)), leading to concrete realizations
of non-singular classical bounces. Here we also find
that EdGB gravity can violate the NCC in spherical gravitational
collapse. 

Also as we find here, for some Horndeski
theories in cosmological settings the equations appear to have
regimes where mixed type character is present. However, as
far as we are aware, 
all these analysis have been carried out at the level of linear perturbations 
about a cosmological background solution, and often the corresponding
elliptic regions are ascribed to be subject to a {\em gradient}, or {\em Laplace
instability}. This is a misnomer in a sense, as the ``instability''
is an artifact of analyzing an elliptic region of a partial differential
equation (PDE) assuming it were
hyperbolic (as opposed to a physical instability in a system
described by hyperbolic PDEs where, for example,
exponentially growing modes can be excited). 
As in our case, this means those Horndeski scenarios do not admit a well-posed
hyperbolic IVP, but does not imply that a sensible
interpretation as a mixed type problem is impossible. 

Note that when many of these modified gravity theories are
directly applied
to address questions on a cosmological scale, things can ``break'' on 
smaller scales such as the solar system, compact objects etc., and
vice versa (hence the need to invent screening or ``chameleon'' mechanisms).
This is certainly the case with EdGB gravity, and so for it to
have a chance of still being interesting and viable for BH mergers
(and barring invention of a screening mechanism)
we must assume the EdGB scalar field is irrelevant on cosmological scales. 
If it did have a large cosmological value (where large means
its contribution to the normalized energy density of the universe
is $\Omega_{EdGB}\sim O(1)$), to avoid formation
of elliptic regions on smaller scales and subsequent breakdown of the
IVP would require
a coupling parameter so small it would be completely uninteresting
for GW tests of GR~\cite{PhysRevD.99.084014}. Moreover, even if one
assumed such mixed type character was benign and would not
lead to unexplained phenomena on smaller scales, measurement
of the speed of GWs implied by the binary neutron star merger GW170817
together with counterpart
electromagnetic signals~\cite{TheLIGOScientific:2017qsa}
rules out large couplings
if there are cosmologically relevant scalars~\cite{Tattersall:2018map}.

Mixed type behavior and elliptic region formation has been observed
and discussed in the context of collapse simulations
of other modified gravity theories
\cite{
Akhoury:2011hr,Leonard:2011ce,Brito:2014ifa,Bernard:2019fjb}.
Reference \cite{Bojowald:2015gra} discusses the
appearance of mixed type PDEs in 
loop quantum gravity models of the early universe,
and in the Hartle-Hawking no boundary
proposal. Interestingly, there the signature change
is interpreted as a property of the model, rather than signalling
a pathology, and proposals are made to solve the corresponding
mixed type equations. Reference \cite{Stewart_2001} provides a similar
discussion of mixed type problems, and proposes methods
to solve this class of PDE in the context of numerical relativity.
A more complete account of the appearance 
of mixed type PDEs in physics and applied
mathematics, and some of the attempts to systematically understand
them, may be found in \cite{otway2015elliptic}.
\subsection{\label{sec:conventions}
	Layout of the remainder of the paper, and conventions}

An outline for the remainder of the paper is as follows.
In Section \ref{sec:basic_equations}, we describe the particular variant
of EdGB gravity we study, write out the form of the equations
within the spherically symmetric ansatz, discuss relevant
initial, boundary and regularity conditions, and briefly
mention the numerical methods we use to solve these
equations (more details on the numerics are given in
\ref{sec:numerical_methods}).
We then describe the main analysis tools we employ to understand
properties of the solutions : the characteristics of the
theory in Section \ref{sec:hyperbolicity_analysis},
a quasi-local mass measure in Section \ref{sec:quasi_local_mass},
and the NCC
in Section \ref{sec:null_convergence_condition}. 
Following that we give results from numerical solutions
of several representative members of our initial data family : Section  
\ref{sec:weak_field_weak_coupling} contains a case
in the weak field, weak coupling regime, Section
\ref{sec:strong_field_weak_coupling} contains 
a case from the strong field, weak coupling regime,
and Section \ref{sec:weak_field_strong_coupling} discusses
several cases from the (moderately) weak field, strong coupling
regime (this was the regime initially presented in~\cite{PhysRevD.99.084014},
where we also give results scaling to the truly weak field,
strong coupling limit).
We discuss potential future directions in the conclusion; in particular
to study long term BH stability in the weak coupling
regime, or early time behavior in the strong field, strong coupling
regime will require the use of horizon penetrating coordinates.

We give some details of the derivation of the EdGB equations in
\ref{sec:EdGB_eom}, the specific form of the components of
the tensor equations of motion within our spherically symmetric
ansatz in \ref{sec:EdGB_equations_of_motion}, a second 
method to compute the
characteristics in \ref{sec:characteristics_method_two}
(largely equivalent to the method described in Section \ref{sec:hyperbolicity_analysis})
a derivation of the
`decoupled' EdGB scalar profile about a Schwarzschild black hole
background in \ref{sec:static_decoupled}, and a description
of all the numerical methods we employed to solve the EdGB PDEs in 
\ref{sec:numerical_methods}.

We used geometrized units where $G=c=1$, and use MTW \cite{misner1973gravitation} 
sign conventions for the metric tensor, etc.

\section{\label{sec:basic_equations}Basic equations}
\subsection{\label{sec:EdGB_gravity}
	Shift-symmetric dilaton Gauss-Bonnet gravity}
	The action for the EdGB gravity theory we consider is 
\begin{equation}
\label{eq:EdGBAction}
	S = \frac{1}{2}\int d^4x\sqrt{-g}
	\left(R - (\nabla\phi)^2 + 2\lambda\phi\mathcal{G}\right)
	,
\end{equation}
	where $R$ is the Ricci scalar, $g$ is the determinant of the
        metric tensor $g_{\mu \nu}$, $\phi$ is the dilaton field, and
        $\mathcal{G}$ is the Gauss-Bonnet scalar that can
        be written in terms of the Riemann tensor $R_{\rho\sigma\mu\nu}$ as
\begin{equation}
\label{eq:GaussBonnetDefinition}
	\mathcal{G}
	\equiv
	\frac{1}{4}\delta^{\mu\nu\alpha\beta}_{\rho\sigma\gamma\delta}
	R^{\rho\sigma}{}_{\mu\nu}R^{\gamma\delta}{}_{\alpha\beta}
	,
\end{equation}
	with $\delta^{\mu\nu\alpha\beta}_{\rho\sigma\gamma\delta}$ the
generalized Kronecker delta. In our units the Gauss-Bonnet coupling constant $\lambda$
has dimension $[L]^{2}$. Varying \eref{eq:EdGBAction} with respect to
$g^{\mu\nu}$ and $\phi$ (see \ref{sec:EdGB_eom}) we obtain
\numparts
\begin{eqnarray}
\label{eq:EdGB_tensor_eom}
  E^{(g)}_{\mu\nu}&\equiv&
	R_{\mu\nu} - \frac{1}{2}g_{\mu\nu}R
	+ 2\lambda\delta^{\gamma\delta\kappa\epsilon}_{\alpha\beta\rho\sigma}
	R^{\rho\sigma}{}_{\kappa\epsilon}
	\left(\nabla^{\alpha}\nabla_{\gamma}\phi\right)
	\delta^{\beta}{}_{(\mu}g_{\nu)\delta} \nonumber \\
	&-& \nabla_{\mu}\phi\nabla_{\nu}\phi 
	+ \frac{1}{2}g_{\mu\nu}\left(\nabla\phi\right)^2
	= 0 
	, \\
\label{eq:EdGB_scalar_eom}
	E^{(\phi)}&\equiv&
	\nabla_{\mu}\nabla^{\mu}\phi + \lambda\mathcal{G} 
	= 0
	.
\end{eqnarray}
\endnumparts

	There are several theories which go under the name
of EdGB gravity, each of which
differs by the functional form of the coupling between the dilaton
and the Gauss-Bonnet scalar, or the presence of a potential
$V(\phi)$ for the dilaton in the action (we consider $V(\phi)=0$). 
For example, one variant of dilaton Gauss-Bonnet gravity
appears as a leading order term 
in the low-energy effective action to certain string theories
(e.g. \cite{Zwiebach:1985uq,Gross:1986mw}): there the coupling goes as
$\alpha e^{-\gamma\phi}$, where $\alpha$ and $\gamma$ are constants
that are set by the string theory in question.  
The theory $(\ref{eq:EdGBAction})$ we consider is equivalent to this 
to leading order in the dilaton coupling (with $\alpha\gamma \propto \lambda$,
and recalling that any constant times $\mathcal{G}$ in the action
in 4-dimensional spacetime can be replaced by a boundary term
that does not affect the equations of motion), and goes by several names:
`shift symmetric Truncated Einstein
dilaton Gauss-Bonnet gravity'
(e.g. \cite{Maeda:2009uy,Yagi:2015oca})
or `shift symmetric
Einstein dilaton Gauss-Bonnet gravity' (e.g. \cite{Witek:2018dmd}); for
brevity we will refer to it simply as EdGB gravity.

	EdGB gravity is invariant under $\phi\to-\phi$ and $\lambda\to-\lambda$.
In this work we present simulations with initial data for $\phi$ that is
everywhere positive, and consider $\lambda$ with positive and negative
values.
\subsection{\label{sec:EdGB_gravity}Spherical symmetry and coordinate ansatz}
	We work in spherical symmetry in polar coordinates, and use the following ansatz for the line element:
\begin{equation}
\label{eq:line_element_polar_coordinates}
	ds^2 
	=
	-e^{2A(t,r)}dt^2 + e^{2B(t,r)}dr^2 
	+ r^2\left(d\vartheta^2 + \mathrm{sin}^2\vartheta d\varphi^2\right)
	.
\end{equation}
	To reduce the dilaton equation of motion (\ref{eq:EdGB_scalar_eom}) to a set of first order 
        PDEs, we define the variables 
\numparts
\begin{eqnarray}
\label{eq:PQDef:Q}	
	Q
	\equiv &
	\partial_r\phi
	, \\
\label{eq:PQDef:P}	
	P
	\equiv &
	e^{-A+B}\partial_t\phi
	.
\end{eqnarray}
\endnumparts
	The equations for the metric (\ref{eq:EdGB_tensor_eom}) and 
scalar (\ref{eq:EdGB_scalar_eom}) retain a similar
structure to the Einstein-Klein-Gordon equations ($\lambda\rightarrow 0$) 
in spherical symmetry; namely, there are no gravitational degrees of
freedom, and all dynamics are driven by the scalar field. Hence, 
we can consider a {\em fully constrained} evolution scheme to solve
for the coupled system of equations. In such a scheme, the metric
variables are solved for using what are essentially 
the elliptic Hamiltonian and momentum constraint equations in GR,
and the scalar equation is treated as hyperbolic. The Klein-Gordon equation is always
hyperbolic in GR (away from coordinate or geometric singularities),
though of course one of the main results of our study
(first described in~\cite{PhysRevD.99.084014}) is this property fails to generically
hold in the strong-coupling limit of EdGB gravity.
Also akin to the Einstein equations, the EdGB system of PDEs
permit a {\em partially constrained} evolution scheme, where a hyperbolic
evolution equation is used for the metric variable $B$ instead of
a constraint equation; as mentioned in the appendix, we have also implemented such a scheme,
and verified we obtain consistent results to within truncation error, though
for brevity all the characteristic analysis and simulation results presented
here pertain to the fully constrained scheme.

        Following the fully constrained evolution strategy,
we take particular algebraic combinations of the nontrivial
components of the EdGB equations of motion (see
\ref{sec:EdGB_equations_of_motion}) to give the following
closed system of PDEs that we solve using numerical methods:
\numparts
\begin{eqnarray}
\label{eq:EdGB_solved:rDer_A}
\fl E^{(A)}  \equiv &\Bigg\{&
		\mathcal{I}^2
	- 	32\lambda^2\mathcal{B}^2
	 +  	128 \lambda^2e^{-2B}\mathcal{B}
		\left(
		1-2\lambda\left(3e^{-2B}+1\right)\frac{Q}{r}
		\right)\frac{\partial_rB}{r} 
        \nonumber \\
\fl
	& & + 256\lambda^3\mathcal{B}^2\left(
		e^{-2B}\partial_rQ 
		- e^{-B} r P\mathcal{K}
		\right)
	\Bigg\}\partial_rA 
	\nonumber \\
\fl
	&+& 4 \lambda e^{-3B}\mathcal{B}
	\Big(
		128\lambda^2e^{2B}r\mathcal{B}P \mathcal{K}
	- 	4 \lambda e^B P^2
	+	e^B\left(r e^{2B}-12\lambda Q\right)Q
	\Big)\partial_rB
        \nonumber \\
\fl
	&-& 512\lambda^3r e^{-B}\mathcal{B}^2
		\mathcal{K}\partial_rP
	- 4\lambda r\mathcal{B}\mathcal{I}
		\partial_rQ
	- \frac{r\mathcal{B}}{2}
		\left(e^{2B}+128\lambda^2\mathcal{K}^2\right)
	\nonumber \\
	&+& 4 \lambda\mathcal{B}\left(-1+128\lambda^2\mathcal{K}^2\right)Q
	+ 2\lambda e^{-2B}Q^3 
	\nonumber \\
\fl
	&+& \left(
		64\lambda^2e^{-2B}r\mathcal{B}
	-	16r^3\lambda^2\mathcal{B}^2-\frac{r^3}{4}
	\right)\left(\frac{Q}{r}\right)^2
	\nonumber \\
\fl
	&+& 4 \lambda r^2e^BP\mathcal{I}\mathcal{B}
		\mathcal{K}
	+ \left(
		16\lambda^2r \mathcal{B}^2
	-	\frac{r}{4}\mathcal{I}
	\right) P^2
	= 0 
	, \\
\label{eq:EdGB_solved:rDer_B}
\fl E^{(B)} \equiv &\Bigg(&1 + 4\lambda\left(1-3e^{-2B}\right)\frac{Q}{r}\Bigg)\partial_rB
	\nonumber \\
\fl
	&-& \frac{r}{4}\left(Q^2+P^2\right)-\frac{1-e^{2B}}{2r}
	+ 4\lambda r\mathcal{B}\left(
	- \partial_rQ
	+ r e^BP \mathcal{K}
	\right)
	=
	0
	, 
	\\
\label{eq:EdGB_solved:tDer_Q}
\fl E^{(Q)} \equiv & &\partial_tQ - \partial_r\left(e^{A-B}P\right)
	= 0 
	, 
	\\
\label{eq:EdGB_solved:tDer_P}
\fl E^{(P)}\equiv&\Bigg(&
	\mathcal{I}
	+ 64\lambda^2e^{-2B}\mathcal{B}\frac{\partial_rB}{r} 
	\Bigg)
	\partial_tP 
	- \left(\mathcal{I} 
	-64\lambda^2e^{-2B}\mathcal{B}\frac{\partial_rA}{r}
	\right)
	\frac{1}{r^2}\partial_r\left(r^2e^{A-B}Q\right)
        \nonumber \\
\fl
	&+& 
		16\lambda e^{A-B}\mathcal{I}
		\left(
		\frac{\partial_rA}{r}\frac{\partial_rB}{r} - \mathcal{K}^2
		\right)
	+ 4\lambda e^{A-B}\mathcal{B}\Bigg[
	  \left(P^2-Q^2\right)
	+ 32\lambda r Q \mathcal{K}^2
	\nonumber \\
\fl
	&-& 16 \lambda e^{-2B}\frac{Q}{r} (\partial_rA)^2
	+ 16 \lambda e^{-B} \left(
			\left(\partial_rB-\partial_rA\right)P - 2\partial_rP
		\right) \mathcal{K}
	+ 2\frac{\partial_rB}{r}
	\nonumber \\
\fl
	&+& 2 \left\{
			-1 - 16\lambda e^{-2B}\frac{Q}{r}
			-2r\left(1-4\lambda e^{-2B}\frac{Q}{r}\right)\partial_rB
		\right\}\frac{\partial_rA}{r}
	\Bigg]	
	= 0 
	, 
\end{eqnarray}
\endnumparts	
	where 
\begin{eqnarray}
\mathcal{B}\equiv (1-e^{-2B})/r^2,\ \ 
\mathcal{I}\equiv 1 - 8\lambda e^{-2B}Q/r, \ {\rm and} \nonumber \\
	\mathcal{K} \equiv
	e^B\frac{
		\frac{PQ}{2} 
		+ 4 \lambda \mathcal{B}\left(-P\partial_rB+\partial_rP\right)
	}{
	e^{2B} + 4\lambda\left(-3 + e^{2B}\right)\frac{Q}{r}
	}
	.
\end{eqnarray}	
In particular
(\ref{eq:EdGB_solved:tDer_Q}) and (\ref{eq:EdGB_solved:tDer_P}) define
the PDE evolution equations for $Q$ and $P$ respectively;
(\ref{eq:EdGB_solved:rDer_A}) and (\ref{eq:EdGB_solved:rDer_B}) contain
no time derivatives and are the ODE (ordinary differential equation) 
constraint equations for $A$ and $B$ respectively.
\subsection{\label{sec:boundary_conditions} Boundary and regularity conditions}
We discretize the above equations over a domain $r\in[0..r_{\rm max}]$. 
At the origin $r=0$ we require regularity of the fields, leading to
\numparts
\begin{eqnarray}
	\partial_rA(t,r)\Big|_{r=0} 
	& = 0
	, \\
	B(t,r)\Big|_{r=0} = 0, \ \ \ \partial_rB\Big|_{r=0}
        & = 0
        , \\
	Q(t,r)\Big|_{r=0} 
	& = 0
	, \\
	\partial_rP(t,r)\Big|_{r=0} 
	& = 0
	.
\end{eqnarray}
\endnumparts
Equations (\ref{eq:EdGB_solved:rDer_A},\ref{eq:EdGB_solved:rDer_B}) are first
order ODEs for $A$ and $B$, so strictly speaking we can only impose
one boundary condition at one of the boundaries for each. In practice
we integrate from $r=0$ to $r_{\rm max}$, setting $A(t,r=0)=0$
and $B(r,r=0)=0$; with the scalar field variables $P$ and $Q$ appropriately regular
as above, the structure of the field equations guarantees that $A$ and $B$
also satisfy the above regularity conditions.
Our coordinate system (\ref{eq:line_element_polar_coordinates}) has residual gauge 
freedom in that we can rescale $t$ by an arbitrary function of itself, and
we use this to rescale $A(t,r)$ after each ODE integration step so
that $A(t,r)|_{r=r_{max}}=0$. In that way our time coordinate $t$
measures proper time of static observers at the outer boundary.

For $Q$ and $P$ at the outer boundary we impose the following approximate
outgoing radiation boundary conditions:
\numparts
\begin{eqnarray}
	\partial_tQ + \frac{1}{r}\partial_r\left(rQ\right)
	\Big|_{r=r_{\max}} 
	&= 0, \\
	\partial_tP + \frac{1}{r}\partial_r\left(rP\right)
	\Big|_{r=r_{\max}} 
	&= 0.
\end{eqnarray}
\endnumparts

\subsection{\label{sec:initial_data}
	Initial data}
For initial data, we are free to choose $P(t=0,r)$ and $Q(t=0,r)$ (subject to
the regularity conditions described in the previous subsection).
For the simulation results presented here, we begin with the
following family of initial data for $\phi(t=0,r$):
\begin{equation}
\label{eq:initial_data_family_1}
	\phi(t,r)\Big|_{t=0}
	=	
	a_0 \left(\frac{r}{w_0}\right)^2
	\mathrm{exp}\left(-\left(\frac{r-r_0}{w_0}\right)^2\right)
	,
\end{equation}
	where $a_0$, $w_0$, and $r_0$ are constants. This then gives
$Q(t,r)|_{t=0}=\partial_r\phi|_{t=0}$, and we choose $P$
so that the scalar pulse is initially approximately
ingoing: 
\begin{equation}
	P(t,r)\Big|_{t=0}
	=
	- \frac{1}{r}\phi(t,r) - Q(t,r)
	\Big|_{t=0}
	.
\end{equation}
Because of spherical symmetry and our constrained evolution scheme,
the only ``free'' data for the metric variables $A$ and $B$ is the
overall scale of $A$, which as discussed in the previous subsection
we set so that $t$ measures proper time for static observers at
the outer boundary of our domain.

\subsection{\label{sec:numerical_methods_brief} Numerical solution methods in brief} 
We numerically solve 
Equations \eref{eq:EdGB_solved:rDer_A}-\eref{eq:EdGB_solved:tDer_P} using
(overall) second order accurate finite difference techniques. We implemented several
different solution methods as detailed in ~\ref{sec:numerical_methods}.
In particular, to gain confidence that the late time 
convergence problems for strong coupling cases are due to the character of the $(P,Q)$ subsystem
changing from hyperbolic to elliptic in a certain region, and not due to 
an unstable numerical evolution method, we explored 
two completely different discretization and evolution schemes
for $(P,Q)$ : (i) a Crank-Nicolson method (with both a Newton-Gauss-Seidel iterative solver,
and Newton iteration together with a fully implicit matrix inversion for each
linear step of the Newton iteration), and (ii) a fourth order in time Runge-Kutta (method of lines)
solver. We integrate the constraint ODEs \eref{eq:EdGB_solved:rDer_A} and 
\eref{eq:EdGB_solved:rDer_B} with a trapezoidal method, which is `A stable'~\cite{atkinson2011numerical}
(we also experimented with a few of variants to deal with the non-linearities
in the equation for $B$, 
as outlined in the appendix). All schemes give solutions consistent with each
other to within truncation error. Here then, all results we show were obtained
using the iterative Crank-Nicholson scheme. We further experimented
with a range of Courant-Friedrichs-Lewy (CFL) factors $0.01-0.5$, confirming
results do not qualitatively depend on this; the simulations discussed here 
used a CFL factor of $0.25$.
\section{\label{sec:hyperbolicity_analysis}Hyperbolicity analysis}
	We briefly summarize the theory of characteristics; standard references
include \cite{courant1962methods,whitham2011linear,kreiss1989initial}.
Consider a system of first order PDEs
\footnote{Through field redefinitions essentially any
system of PDEs may be written in this form.}
\begin{equation}
\label{eq:system_first_order_pde}
	E^I\left(v^J,\partial_av^K\right) = 0
	,
\end{equation} 
	where $I,J,K$ index the $N$ equations of motion and dynamical fields $v^J$,
and $a$ indexes the $n$ coordinates $\{x^a\}$ of the underlying (spacetime)
manifold $M$ (and here because of our restriction to spherical
symmetry $a$ only runs over the $(t,r)$ coordinates).
The \emph{principal symbol} is defined to be
\begin{equation}
	\mathfrak{p}^I_J\left(\xi_a\right)
	\equiv \frac{\delta E^I}{\delta(\partial_av^J)}\xi_a
	,
\end{equation}
	where $\xi_a$ is an $n$ dimensional covector. A \emph{characteristic
surface} $\Sigma\subset M$ by definition
satisfies the \emph{characteristic equation}
\begin{equation}
\label{eq:characteristic_equation}
	\mathrm{det}\left(\mathfrak{p}^I_J\left(\partial_a\Sigma\right)\right)
	= 0
	,
\end{equation}
	and \eref{eq:characteristic_equation} is the \emph{eikonal equation}
for the characteristic surface. 

	For a physical interpretation of characteristics,
we consider a system of $N$
first order PDEs for $N$ fields (i.e. of the form
\eref{eq:system_first_order_pde})
that is \emph{totally hyperbolic}: i.e it has
$N$ real (possibly degenerate)
characteristic surfaces\footnote{For a system that is not totally
hyperbolic we could instead consider a totally hyperbolic subsystem;
see e.g. Section \ref{sec:characteristics_method_one}. Our treatment
of characteristics roughly follows that
of \cite{christodoulou2008mathematical}.}.
Consider the solution to small amplitude high frequency wave solutions:
$v_0^I e^{ik_ax^a/\epsilon}$, with $0<\epsilon\ll1$.
Solutions of this form to leading order in $\epsilon$
satisfy $\mathfrak{p}^I_J(k_a)v_0^J = 0$. Nontrivial solutions to
this equation exist if and only if
$\mathrm{det}\left(\mathfrak{p}^I_J(k_a)\right)=0$;
i.e. if and only if the wave vector satisfies 
\eref{eq:characteristic_equation}.
Thus the wavefronts of high frequency wave solutions propagate on the
characteristic surfaces. The characteristic surfaces
locally delimit the causal region of influence for hyperbolic PDE
(e.g. \cite{Geroch:1996kg}). 

	In local coordinates, letting $t$
index the timelike coordinate and $i$ index the spacelike
coordinates of the background geometry, the speed of
these perturbations for the $n^{th}$ characteristic
is given by $c^{(n)}=(v^{(n)})^i/v^t$, where $(v^{(n)})^{\mu}$ is
a vector parallel to the $n^{th}$ characteristic surface. We may relate
$c^{(n)}$ to the characteristic covector by noting that since
locally the $n^{th}$ characteristic covector is equal to the gradient
of the $n^{th}$ characteristic surface,
$\xi^{(n)}_{\mu}=\partial_{\mu}\Sigma^{(n)}$, then  
$(v^{(n)})^{\mu}\xi^{(n)}_{\mu}=0$, from which we find 
$c^{(n)}=-\xi^{(n)}_t/\xi^{(n)}_i$. For a simple example of this
procedure, consider 
the $1+1$ dimensional scalar transport equation
$\partial_t\psi + v\partial_x\psi=0$.
The symbol is $\mathfrak{p}=\xi_t+v \xi_x$, the characteristic
equation is $\xi_t+v\xi_x=0$, 
and the speed of propagation
along the characteristic is  $-\xi_t/\xi_x=v$.

	We compute the characteristic vectors and speeds
for the system of PDEs 
\eref{eq:EdGB_solved:rDer_A},
\eref{eq:EdGB_solved:rDer_B},
\eref{eq:EdGB_solved:tDer_Q}, and
\eref{eq:EdGB_solved:tDer_P} in two different ways. In the first, 
discussed below, we only consider the $P,Q$ evolution subsystem,
eliminating all $A$ and $B$ gradients from these equations
using the constraints. In the second (discussed in \ref{sec:characteristics_method_two}),
which is more for a consistency check than anything else, we apply the characteristic analysis 
verbatim to the full system of equations, obtaining the same
results for $P,Q$ as the first, and confirming that $A$ and $B$ are elliptic.

\subsection{\label{sec:characteristics_method_one}
	Characteristics calculation}
Eliminating $\partial_rA$ and $\partial_rB$ from
Equations \eref{eq:EdGB_solved:tDer_Q} and \eref{eq:EdGB_solved:tDer_P}
using 
Equations \eref{eq:EdGB_solved:rDer_A} and \eref{eq:EdGB_solved:rDer_B},
we write the scalar field system in the same form as before, 
\begin{equation}
\label{eq:totally_hyperbolic_subsystem_QP}
	\tilde{E}^I\left(v^J,\partial_av^K\right)
	= 0
	,
\end{equation}
	but now $I,J,K$ only index the fields $Q$ and $P$. The principal symbol
then reads
\begin{equation}\label{eq:principal_symbol_PQ}
	\mathfrak{p}(\xi) 
	=
	\tilde{\mathfrak{a}} \xi_t
	+ 
	\tilde{\mathfrak{b}} \xi_r
	,
\end{equation}
	where
\numparts
\begin{eqnarray}
	\tilde{\mathfrak{a}}
	\equiv &
	\pmatrix{%
		\delta \tilde{E}^{(Q)}/\delta\left(\partial_tQ\right) & 
		\delta \tilde{E}^{(Q)}/\delta\left(\partial_tP\right) \cr
		\delta \tilde{E}^{(P)}/\delta\left(\partial_tQ\right) &
		\delta \tilde{E}^{(P)}/\delta\left(\partial_tP\right) 
	}
	, \\
	\tilde{\mathfrak{b}}
	\equiv &
	\pmatrix{%
		\delta \tilde{E}^{(Q)}/\delta\left(\partial_rQ\right) & 
		\delta \tilde{E}^{(Q)}/\delta\left(\partial_rP\right) \cr
		\delta \tilde{E}^{(P)}/\delta\left(\partial_rQ\right) &
		\delta \tilde{E}^{(P)}/\delta\left(\partial_rP\right) 
	}
	.
\end{eqnarray}
\endnumparts
	Solving the characteristic equation for the characteristic speeds
$c\equiv-\xi_t/\xi_r$, we obtain
\begin{equation}
\label{eq:characteristic_speeds}
	c_{\pm} =
	\frac{1}{2}
	\left(
		\mathrm{Tr}\left(\tilde{\mathfrak{c}}\right)
		\pm
		\sqrt{
			\mathrm{Tr}\left(\tilde{\mathfrak{c}}\right)^2 
		-	4 \mathrm{Det}\left(\tilde{\mathfrak{c}}\right)
		}
	\right)
	,
\end{equation}
	where
\begin{equation}
	\tilde{\mathfrak{c}} 
	\equiv
	\tilde{\mathfrak{a}}^{-1}\cdot\tilde{\mathfrak{b}}
	,
\end{equation}
	From standard PDE theory, the sign of the discriminant 
\begin{equation}
\label{eq:discriminant_def}
	\mathcal{D}
	\equiv
	\mathrm{Tr}\left(\tilde{\mathfrak{c}}\right)^2
	- 4 \mathrm{Det}\left(\tilde{\mathfrak{c}}\right)
\end{equation}
of \eref{eq:characteristic_speeds} at any point of the
spacetime determines the character of the PDE there:
when $\mathcal{D}>0$ it is hyperbolic,
when $\mathcal{D}=0$ it is parabolic,
and when $\mathcal{D}<0$ it is elliptic.

	In general when $\lambda\neq0$,
$\mathrm{Tr}\left(\mathfrak{c}\right)\neq0$, so
that $c_{+}\neq -c_{-}$. 
In GR ($\lambda=0$), we have
\begin{eqnarray}
	\tilde{\mathfrak{a}}\big|_{\lambda=0}
	= &
	\pmatrix{%
		1 & 0 \cr 0 & 1
	}
	, \\
	\tilde{\mathfrak{b}}\big|_{\lambda=0}
	= &
	\pmatrix{%
		0 & -e^{A-B} \cr -e^{A-B} & 0
	}
	,
\end{eqnarray}
	so that $\mathrm{Tr}\left(\mathfrak{c}\right)=0$, and the characteristic
speeds are $c_{\pm}\big|_{\lambda=0}=\pm e^{A-B}$. The general expressions
for the components of the matrices $\tilde{\mathfrak{a}}$,
$\tilde{\mathfrak{b}}$, and $\tilde{\mathfrak{c}}$
can be obtained through straightforward
algebraic manipulation of
Equations~\eref{eq:EdGB_solved:rDer_A}-\eref{eq:EdGB_solved:tDer_P};
the resultant expressions are long and not particularly insightful, so we do not
write out their full forms here.

\subsection{\label{sec:hyperbolicity_analysis_invariance}
	Invariance of the characteristics under coordinate transformations}
As mentioned, one of the main results of our study is that the EdGB equations
in spherical symmetry can be of mixed elliptic/hyperbolic type in certain scenarios.
Specifically then, evolution, beginning with initial data where the
scalar equation is everywhere hyperbolic, leads to formation of a region where the characteristic structure
switches to elliptic (separated by a parabolic so-called {\em sonic line}, though generically
is a co-dimension one surface and not a ``line''). The elliptic
region is particularly problematic for the validity of EdGB gravity
as a classically well-posed, predictive modified theory of gravity (see the discussion
in~\cite{PhysRevD.99.084014}), and so we would like to be certain that
our identification of the elliptic region is not somehow a coordinate artifact.
It is well-known that the characteristic structure of a PDE is invariant
under so-called point transformations (essentially coordinate transformations
treating all the dependent variables as scalars), though
it is unclear that this
must hold when solving the PDEs of EdGB gravity as a Cauchy IVP
problem in an arbitrary gauge. The problem is that
the structure and even rank of the principal symbol is unknown until 
the gauge equations have been chosen (in the ADM [Arnowitt-Deser-Misner] language, that
would be the equations governing the lapse $\alpha$ and shift vector $\beta^i$: 
we have effectively chosen the equation for the shift to be the
algebraic condition $\beta^i=0$, and our choice of polar-areal coordinates
in spherically symmetry fixes $\alpha$ to within an overall scale). 

However, at least we can show that the characteristic structure is invariant
under coordinate transformations in the following sense (this is effectively
the point-transformation calculation). In our evolution
scheme we compute the characteristic surfaces $\Sigma$ as outlined above, and find
the corresponding co-vectors $\xi_a = \partial_a \Sigma$. If at a given point
$c\equiv -\xi_t/\xi_r$ is purely real, we know the PDE is hyperbolic
at that point, and information will propagate along the characteristic surface.
This could be superluminal, luminal, or sub-luminal relative to the metric
light-cone depending on whether $\xi^a \xi_a$ is negative, zero, or positive
respectively, but the scalar equation is still hyperbolic and will have
its own causal-cone of influence. If $c$ has an imaginary component this is
no longer true, and the PDE is elliptic. The question then is whether
this property of the characteristic surface is invariant under coordinate 
transformations, and the answer is yes. 
For consider a general coordinate transformation respecting
the spherical symmetry of the spacetime : let $x^a=x^a(\tilde{x}^{\tilde{a}})$
where $x^a$ and $\tilde{x}^{\tilde{a}}$ denote the $(t,r)$ and $(\tilde{t},\tilde{r})$
coordinates respectively, and the Jacobian of the transformation is 
$\Lambda^a{}_{\tilde{a}} \equiv \partial x^a/\partial \tilde{x}^{\tilde{a}}$
(with all coordinates and metrics real). 
Then $\xi_{\tilde{a}} = \xi_a \Lambda^a{}_{\tilde{a}} $, the new
coordinate speed is $\tilde{c}\equiv -\xi_{\tilde{t}}/\xi_{\tilde{r}}$, and
it is straight-forward
to calculate that
\begin{equation}
{\rm Im} (\tilde{c}) = {\rm Im}(c) \frac{\det[\Lambda^a{}_{\tilde{a}}]}{Z Z^\ast},
\end{equation}
where $Z\equiv-c \Lambda^t{}_{\tilde{r}} + \Lambda^r{}_{\tilde{r}}$, and $Z^\ast$
its complex conjugate.
In other words, as long as the transformation is non-singular
an imaginary piece to $c$ in one coordinate system implies one in all.

\subsection{\label{sec:horizons}
	Horizons}
As mentioned above, when hyperbolic, the causal cones of the scalar degree
of freedom
in EdGB gravity ($\lambda\neq 0$) do not generally coincide with those
of the spacetime. The latter
would govern the speed of propagation of fields minimally coupled to
the metric,
such as a massless scalar field or a Maxwell field, and gravitational
waves (which
are not present in spherical symmetry). Regarding metric horizons,
our coordinate system does not allow evolution through
formation of a black hole, as the geometric light speeds are
$c_{g\pm}\big|=\pm e^{A-B}$; i.e. the metric is necessarily
singular along horizons. In strong-field evolutions we estimate
that gravitational collapse occurs when $c_{g+}$ 
starts to evolve to zero at a finite radius; evolution beyond horizon
formation will require the use of horizon penetrating coordinates,
which we leave to a future study.

\section{\label{sec:quasi_local_mass}Quasi-local mass}
	In spherical symmetry in GR, a
standard definition of quasi-local mass is the Misner-Sharp mass
\cite{PhysRev.136.B571,Szabados2009} (sometimes also referred
to as the Hawking-Israel or Hernandez-Misner mass, e.g.~\cite{Abreu:2010ru})
\begin{equation}
\label{eq:misner_sharp_definition}
	m_{MS}(t,r)\equiv
	\frac{r}{2}\left(1-(\nabla r)^2\right) = (r/2)\left(1-e^{-2B(t,r)}\right)
	,
\end{equation}
	where $r$ is the areal radius, and the last term on the
        right is the specific form it takes in our coordinate system.
The Misner-Sharp mass satisfies several useful criteria for
a quasi-local mass (e.g. \cite{Hayward:1994bu}). For
example, in asymptotically flat spacetimes it reduces to the
ADM mass
at spatial infinity and the Bondi-Sachs mass at future null infinity.
A further useful property of the Misner-Sharp mass is in spherical
symmetry one can relate it to the charge associated
with the {\em Kodama current}~\cite{Kodama:1979vn},
which satisfies a conservation law 
purely from properties of the Einstein tensor in spherical
symmetry: one does need to {\em a priori} connect
the Einstein tensor to the matter stress energy tensor
to prove this (see e.g.~\cite{Abreu:2010ru}). 
Therefore it is reasonable to use the Misner-Sharp mass
in spherically symmetric EdGB gravity as a measure of geometric mass. Then
(akin to GR), if desired we can use the EdGB equations of motion
to relate it to an integral of an effective matter energy density.
Specifically, we write the EdGB equations (\ref{eq:EdGB_tensor_eom}) 
as $G_{\mu\nu}=\mathcal{T}_{\mu\nu}$, with
\begin{equation}
\label{eq:stress_energy_EdGB}
	\mathcal{T}_{\mu\nu}
	\equiv
	-2\lambda\delta^{\gamma\delta\kappa\lambda}_{\alpha\beta\rho\sigma}
	R^{\rho\sigma}{}_{\kappa\lambda}
	\left(\nabla^{\alpha}\nabla_{\gamma}\phi\right)
	\delta^{\beta}_{(\mu}g_{\nu)\delta}
	+ \nabla_{\mu}\phi\nabla_{\nu}\phi 
	- \frac{1}{2}g_{\mu\nu}\left(\nabla\phi\right)^2
	.
\end{equation}
Then, replacing $G_{\mu\nu}$ with $\mathcal{T}_{\mu\nu}$ in the Kodama
current (see e.g.\cite{Abreu:2010ru}), a short calculation gives
the following
integral for the Misner-Sharp mass in our coordinate system (assuming regularity
at $r=0$):
\begin{equation}
\label{eq:ms_set_effective}
m_{MS}(t,r)=\frac{1}{2}
	\int_0^rdr^{\prime}(r^{\prime})^2e^{-2A(t,r^{\prime})}\mathcal{T}_{tt}(t,r^{\prime}).
\end{equation}
The effective stress tensor $\mathcal{T}_{\mu\nu}$
does not always satisfy the usual energy conditions, hence $m_{MS}(t,r)$ is not
necessarily a monotonically increasing function of $r$, as it is in GR
coupled to ``ordinary'' matter. We will show some examples below illustrating
the non-monotonicity of $m_{MS}$.

\section{\label{sec:null_convergence_condition}
	Null convergence condition}
	The null convergence condition (NCC) is
\begin{equation}
\label{eq:null_convergence_condition}
	R_{\mu\nu}l^{\mu}l^{\nu}\geq 0
	,
\end{equation}
	for all null vectors $l^{\alpha}$.
        The NCC plays a role
in, for example the classical
black hole and cosmological singularity theorems \cite{hawking1975large},
the laws of black hole mechanics and dynamical horizons \cite{1973CMaPh..31..161B,PhysRevD.49.6467,2004LRR.....7...10A},
and in the ``topological censorship" theorems
\cite{PhysRevLett.71.1486,Galloway:1999bp}.
It is often stated that these theorems and properties
rely on the null energy condition (NEC), $T_{\mu\nu}l^{\mu}l^{\nu}\geq 0$, however that comes
from replacing the Ricci tensor in the above with an equivalent
function of the stress energy tensor using the Einstein equations.
We could like wise recast our analysis in terms of a NEC
using the effective stress energy tensor introduced in the previous
section, though we prefer the geometric interpretation of the NCC.

Related to the fact that the Misner-Sharp mass does not always monotonically
increase with radius as discussed in the previous section, EdGB gravity
does not generically satisfy the NCC (coupling to matter
than satisfies the NEC). This can be seen 
by contracting \eref{eq:EdGB_tensor_eom} with $l^{\mu}l^{\nu}$ to
compute the explicit form of $ R_{\mu\nu}l^{\mu}l^{\nu}$:
\begin{eqnarray}
\label{eq:null_convergence_condition_EdGB}
\fl	R_{\mu\nu}l^{\mu}l^{\nu}
	= 
	\frac{1}{1-4\lambda\nabla_{\alpha}\nabla^{\alpha}\phi}
	\Bigg(
	\left(l^{\mu}\nabla_{\mu}\phi\right)^2
	 &
	+
	2\lambda l^{\mu}l^{\nu}\Big(
		\left(\nabla_{\mu}\nabla_{\nu}\phi\right)R
	\nonumber \\
\fl	& 
	-	4\left(\nabla_{\mu}\nabla_{\alpha}\phi\right)
		R^{\alpha}{}_{\nu}
	-	2\left(\nabla_{\alpha}\nabla_{\beta}\phi\right)
		R^{\alpha}{}_{\mu}{}^{\beta}{}_{\nu}
	\Big)
	\Bigg)
	.
\end{eqnarray}
	Here, the only term that is manifestly positive definite
is the kinetic term of the scalar in the small coupling
($\lambda\rightarrow 0$) limit. We will show examples below
of scenarios where the NCC is violated during evolution in EdGB gravity
(and the regions where it does roughly coincide with negative effective energy
density in the Misner-Sharp mass, and is present where the equations
become elliptic); specifically,
we numerically evaluate $R_{\mu\nu}l^{\mu}l^{\nu}$
for outgoing null vectors $l^{\mu}\equiv(e^{-A},e^{-B},0,0)$.
\section{\label{sec:weak_field_weak_coupling}
	Numerical results: weak field, weak coupling}
\begin{figure}%
    \centering
   	\includegraphics[width=1.0\textwidth]{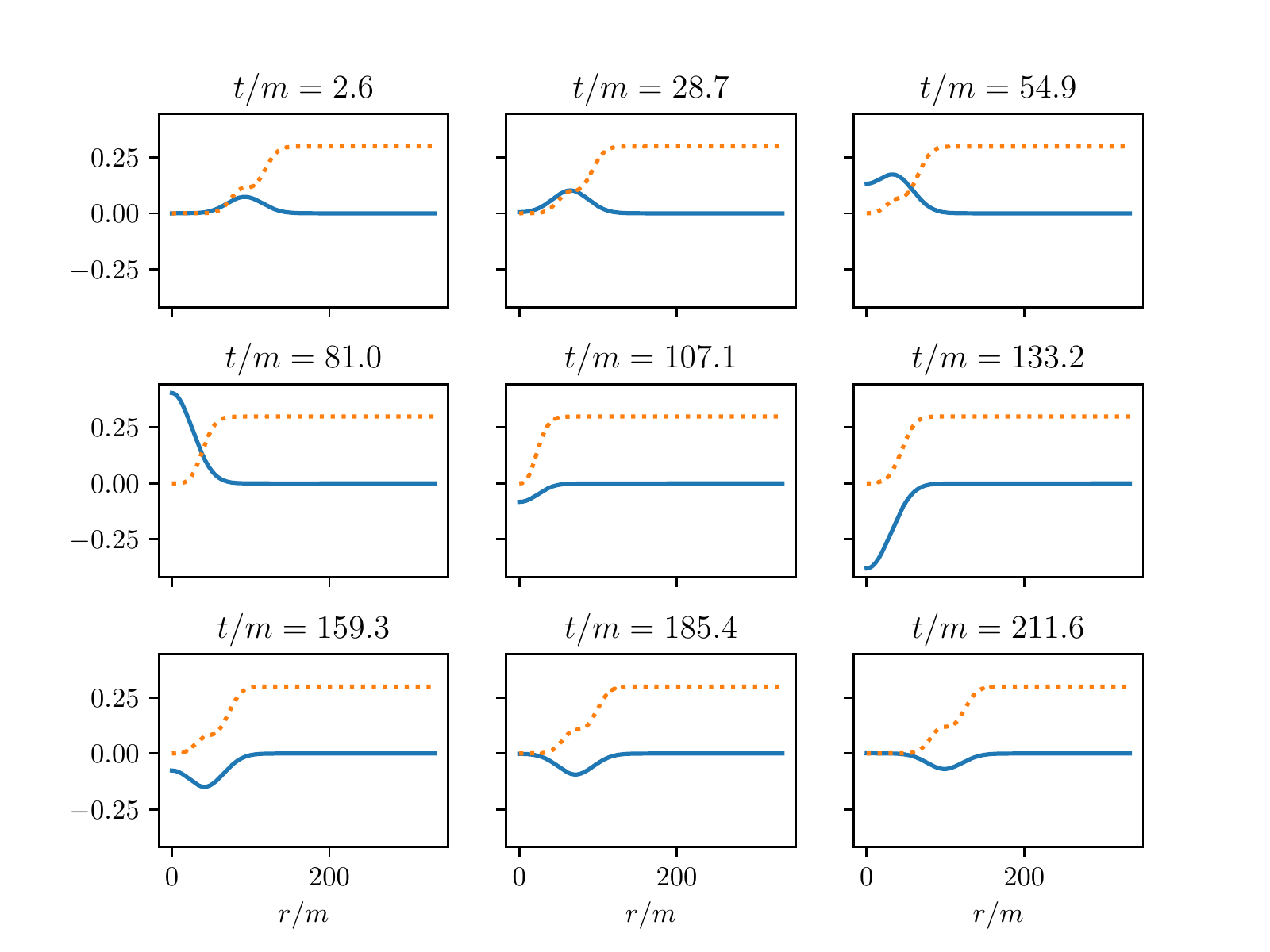}.
	\caption{The scalar field profile $\phi$ (blue solid line) and Misner-Sharp
        mass  $m_{MS}$ (orange dashed line) from a weak field, weak coupling run
        with scalar field initial data parameters (\ref{eq:initial_data_family_1})
	$a_0=0.01$, $r_0=25$, $w_0=10$, and $\lambda=0.1$, $r_{\rm max}=100$
	(discretized with $N_r=2^{12}+1$ points). Here, and in all figures,
        we normalize units with respect to $m\equiv m_{MS}(t=0,r=r_{\rm max})$.
        The metric fields (not shown)
        remain smooth and close to their Minkowksi spacetime values throughout.
	}
    \label{fig:multiple_panels_weak_field_weak_coupling}%
\end{figure}

Figure \ref{fig:multiple_panels_weak_field_weak_coupling} shows results
from the evolution of a representative 
member of the initial data family (\ref{eq:initial_data_family_1}) corresponding
to a weak field and weak coupling case: the compaction of the scalar field $m_{ADM}/w_0\ll 1$, 
and $\lambda/w_0^2\ll1$ respectively. The scalar pulse
bounces smoothly off the origin and disperses to infinity. Throughout the
evolution, the characteristics
remain real and close to the geometric null characteristics.
To within truncation error, the Misner-Sharp mass is monotonically increasing
in $r$ and the NCC is preserved. Qualitatively the evolution matches that
of Einstein massless scalar field evolution ($\lambda=0$). 

\section{\label{sec:strong_field_weak_coupling}
	Numerical results: strong field, weak coupling}
\begin{figure}%
    \centering
   	\includegraphics[width=1.0\textwidth]{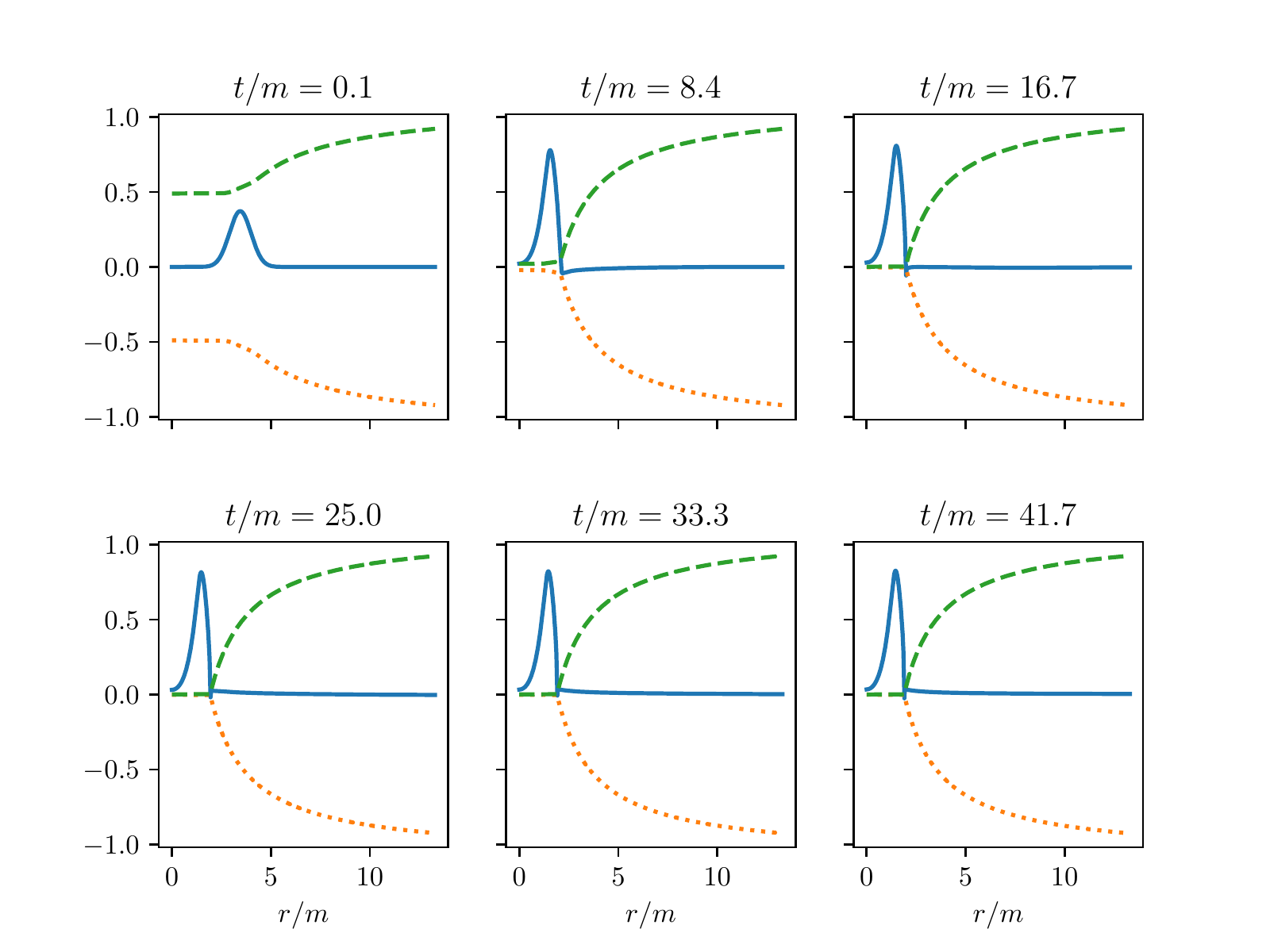}
	\caption{Run with strong field and weak coupling:
	$a_0=0.02$, $w_0=6$, $r_0=25$, $\lambda=1$, $r_{max}=100$, $N_r=2^{12}+1$, and
        $m\sim 7.5$.
        Shown is the scalar field $\phi$ (blue line),
        and corresponding ingoing (orange dots) and
	outgoing (green dashes) characteristic speeds. An apparent
	horizon begins to form soon after evolution begins. That both
        characteristic speeds go to zero inside the horizon $r\lesssim 2m$
        is an artifact of the horizon-avoiding nature of the coordinates,
        as time flow ``freezes'' in this region
        as $A\to-\infty$ here. Outside the horizon the scalar
        field slowly grows, and appears to asymptote to the profile
        expected for a ``hairy'' black hole in EdGB gravity---see Figure~\ref{fig:comparison_scalar_field_to_analytic_value}
        for a zoom-in of the late-time profile (though ``late'' is not
        particularly so in these coordinates, as we quickly loose
        convergence once $A$ and $B$ start to diverge).
	}
    \label{fig:multiple_panels_strong_field_weak_coupling}%
\end{figure}
\begin{figure}%
    \centering
   	\includegraphics[width=0.8\textwidth]{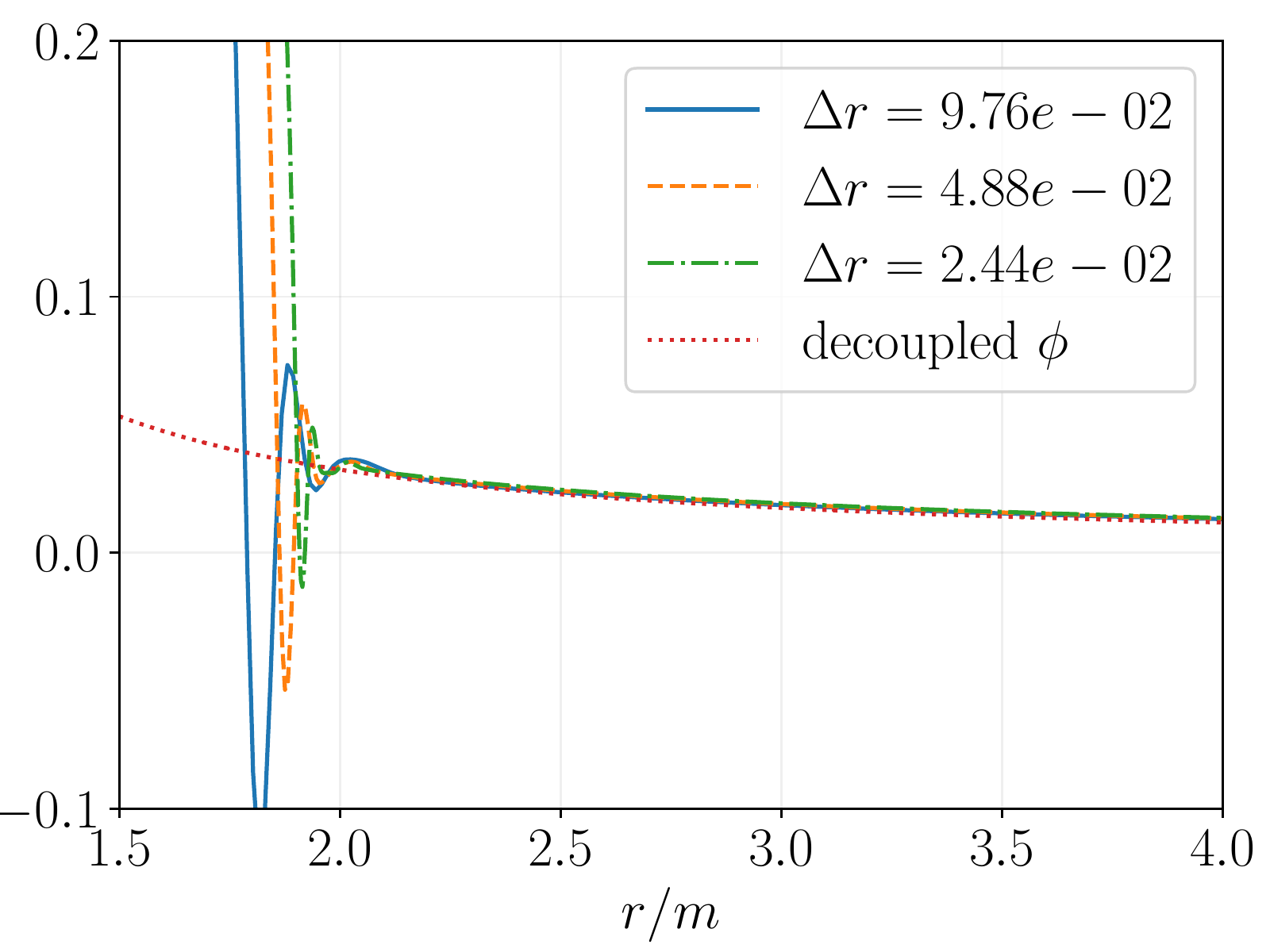}.
	\caption{Scalar field profiles from the evolution of the same
        initial data as shown in Figure ~\ref{fig:multiple_panels_strong_field_weak_coupling}
	($a_0=0.02$, $w_0=6$, $r_0=25$, $\lambda=1$, $r_{max}=100$),
        from simulations with three different resolutions,
        at a single time $t\sim35m$.
        Also shown is the scalar field profile from
        the analytic solution on a static black hole background with the same
	ADM mass as our initial data $m\sim7.5$, computed in the
        decoupling limit (i.e., the scalar field does not back-react
        on the geometry; see
	\ref{sec:static_decoupled}).
	The marked difference in the oscillations in $\phi$ with resolution 
        interior to the horizon $r/m\approx2$ indicates loss of convergence,
	as $A\to-\infty$ here, while $B\to\infty$ 
        on the horizon. However the oscillations do ``converge away'' in the
        sense that evaluated at a fixed time their amplitude decreases
        with increasing resolution. Also, this oscillatory behavior
        does not appear to adversely affect the profile of the field
        outside of the horizon, though a more reliable analysis will require
	horizon penetrating coordinates, which we will implement in a future
	study.
	}
    \label{fig:comparison_scalar_field_to_analytic_value}%
\end{figure}
	We ran simulations with initial data that in GR would form
a geometric horizon at $r\sim r_{gh}$, but weakly perturbed
by EdGB modifications with $\lambda/r_{gh}^2\ll1$. As our 
coordinates are not horizon penetrating,
we cannot evolve the spacetime to or beyond
(geometric) apparent horizon formation,
which in our coordinates
is signaled by $e^{A-B}\to0$ (see Section \ref{sec:horizons}).
At all the finite resolutions we have used there is some
small value of $e^{A-B}$ below which we loose 
convergence (and both metric fields diverge here, with $A\to-\infty$ and $B\to\infty$).
Though before this, we do 
observe that the EdGB scalar field begins to grow near the nascent horizon.
The growth is in qualitative agreement with the conclusions of
\cite{Kanti:1995vq,Sotiriou:2014pfa,Benkel:2016kcq,Benkel:2016rlz,Witek:2018dmd},
where the value of the field at the horizon is expected to asymptote
to a unique value $\phi_{gh}\propto \lambda/r_{gh}^2$ depending upon the mass of the 
black hole (though of course
questions about the ultimate stability of such ``hairy'' black holes, and under what conditions no elliptic regions
form outside the horizon will require numerical solutions using horizon
penetrating coordinates).

Figure~\ref{fig:multiple_panels_strong_field_weak_coupling} shows results of an
example from such a strong field, weak coupling case.
The scalar pulse approaches the origin, then ``freezes" interior to
what will be the eventual horizon
(since the lapse function $\alpha=e^A\to0$ there). Outside, 
the scalar field begins
to grow, and for a while we can follow its evolution 
before the code fails. In Figure \ref{fig:comparison_scalar_field_to_analytic_value} 
we show a zoom-in of the scalar field at such a late time,
together with the expected solution in the decoupling limit
for a regular EdGB scalar field on a static black hole background
(see \ref{sec:static_decoupled}). That this solution
qualitatively matches well with the full nonlinear evolution is
another indication that this is in the weak coupling regime of EdGB gravity.
\section{\label{sec:weak_field_strong_coupling}
	Numerical results: strong coupling, weak field}
We first presented results from the weak field, strong coupling 
regime in~\cite{PhysRevD.99.084014}; here we describe two additional
examples, and give more details. Specifically, we
consider initial data (\ref{eq:initial_data_family_1}) with 
$a_0=0.02$, $r_0=20$, $w_0=8$, and $\lambda=\pm50$; $m\sim 0.93$ for both
cases (so this is fairly compact initial data, but is ``weak'' in the
sense that we are still a factor of a few in mass away from 
initial data that would form a black hole; in~\cite{PhysRevD.99.084014} 
further data was given showing scaling to the truly weak field (low compaction),
strong coupling regime).

\subsection{\label{sec:formation_elliptic_region}
	Characteristics and formation of elliptic regions}
\begin{figure}%
    \centering
    \subfloat[EdGB characteristics]{{\includegraphics[width=0.8\textwidth]{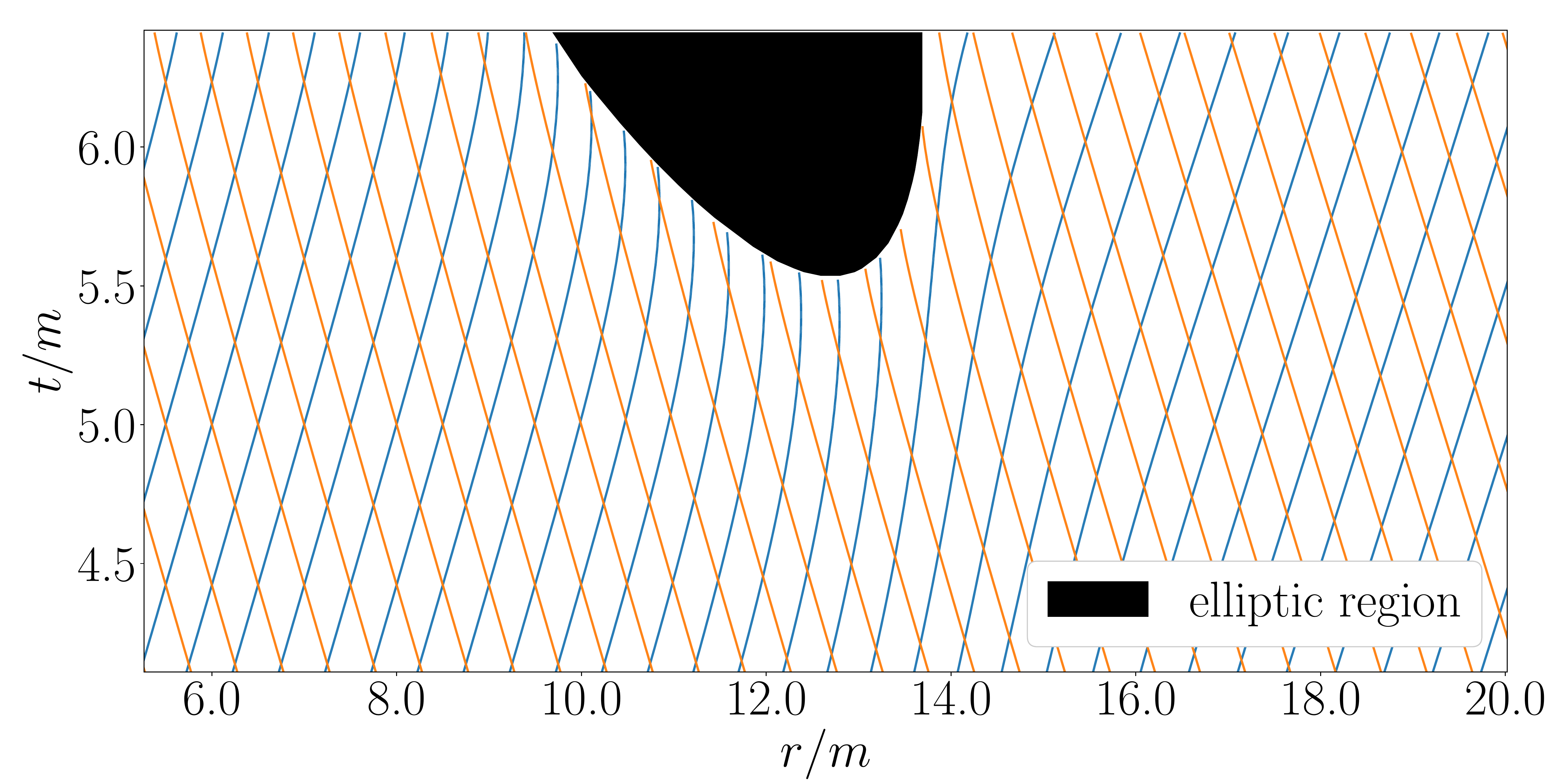} }}
    \\
    \subfloat[Null characteristics]{{\includegraphics[width=0.8\textwidth]{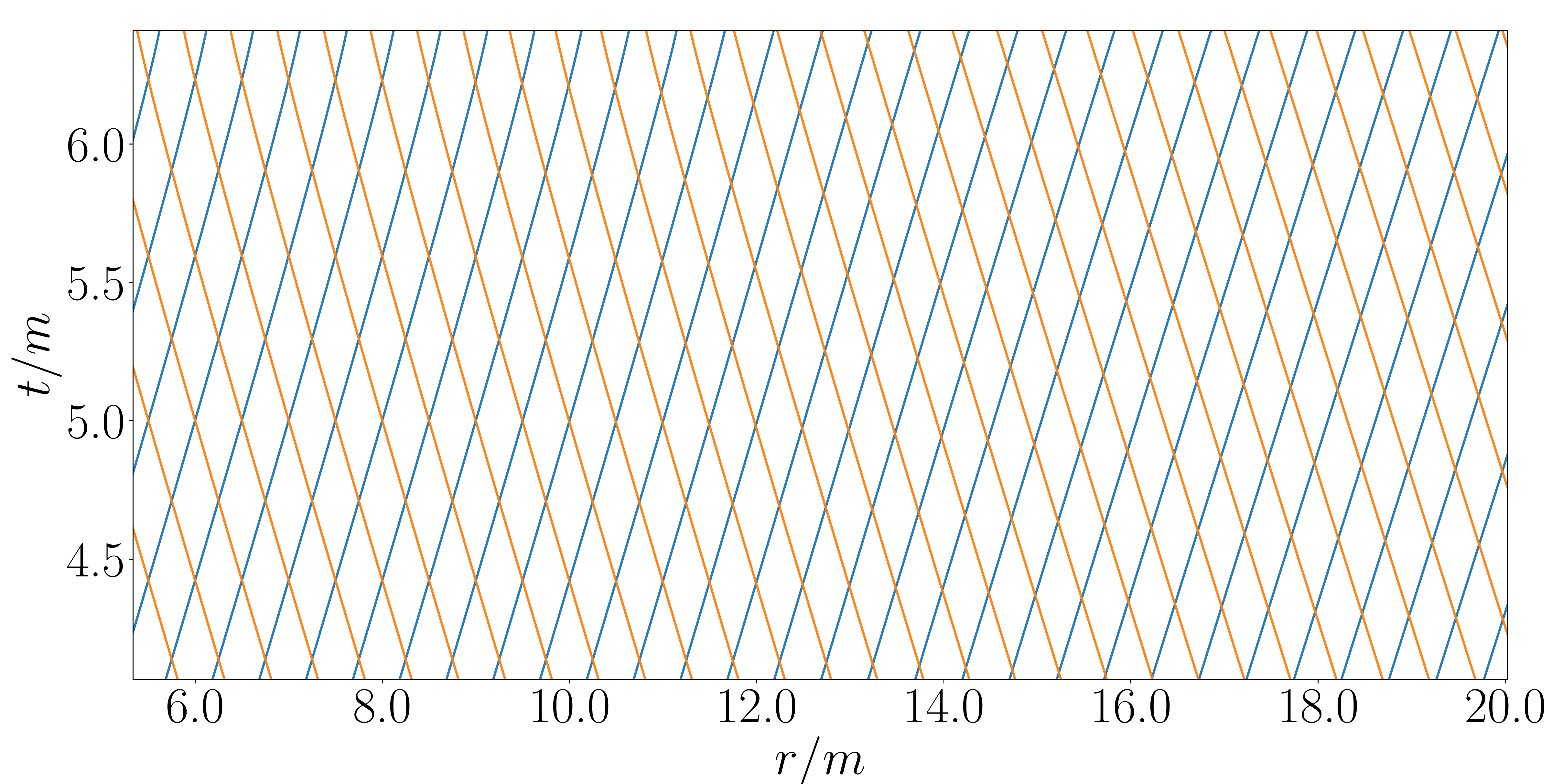} }}
	\caption{
	Characteristic lines from a strong coupling, weak field case:
	$a_0=0.02,w_0=8,r_0=20$, $\lambda=50$, $r_{max}=100$, $N_r=2^{12}+1$; $m\sim 0.93$.
        The top panel shows the characteristics of the principal
        symbol (\ref{eq:principal_symbol_PQ}) of the EdGB equations, the bottom panel
        the spacetime radial null curves. Compare Figure \ref{fig:elliptic_forming_regions_negative_lambda}
        for a case with the same initial data, but opposite sign for $\lambda$.
	}
    \label{fig:elliptic_forming_regions_positive_lambda}%
\end{figure}

\begin{figure}%
    \centering
    \subfloat[EdGB characteristics]{{\includegraphics[width=0.8\textwidth]{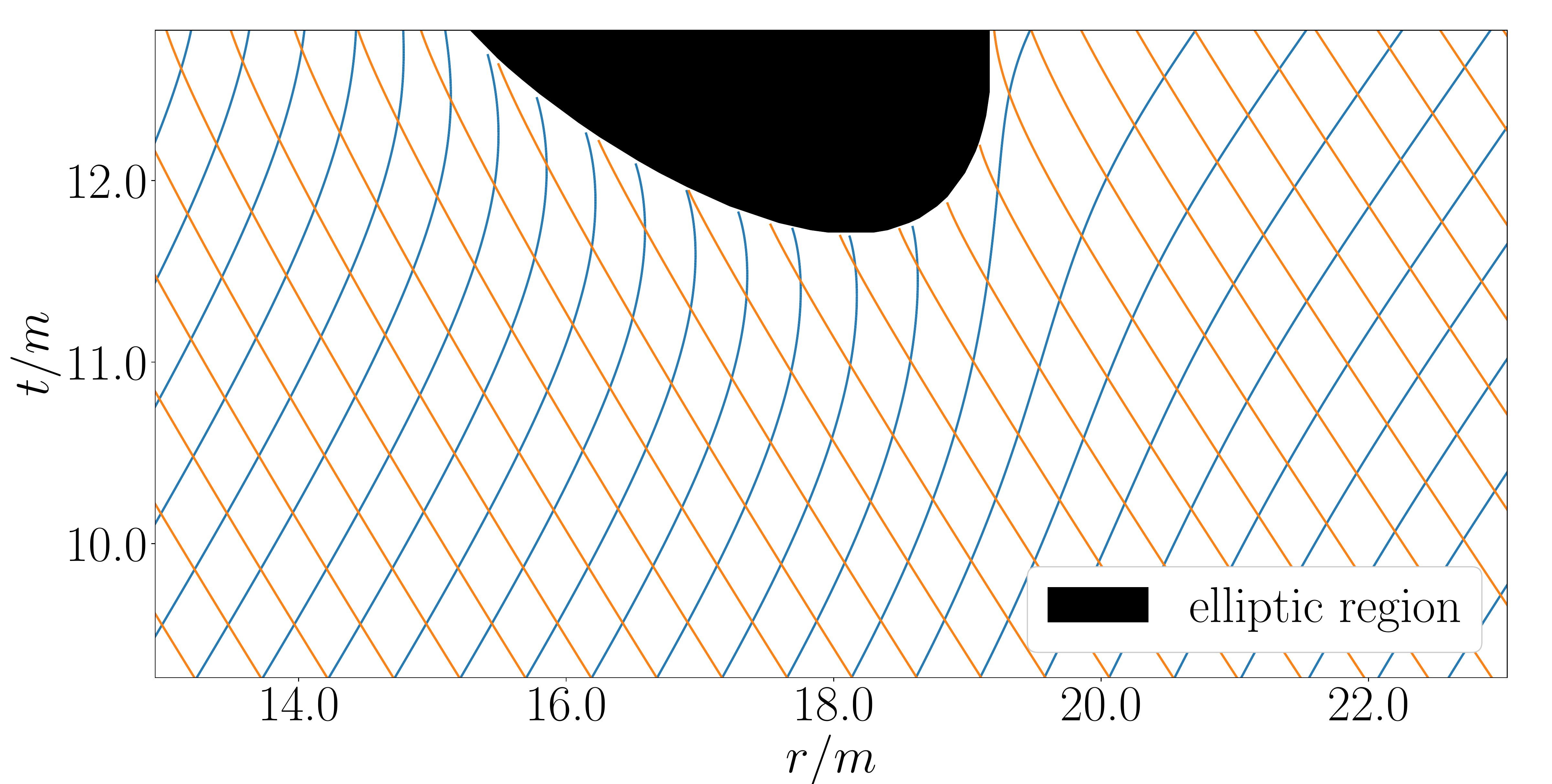} }}
    \\
    \subfloat[Null characteristics]{{\includegraphics[width=0.8\textwidth]{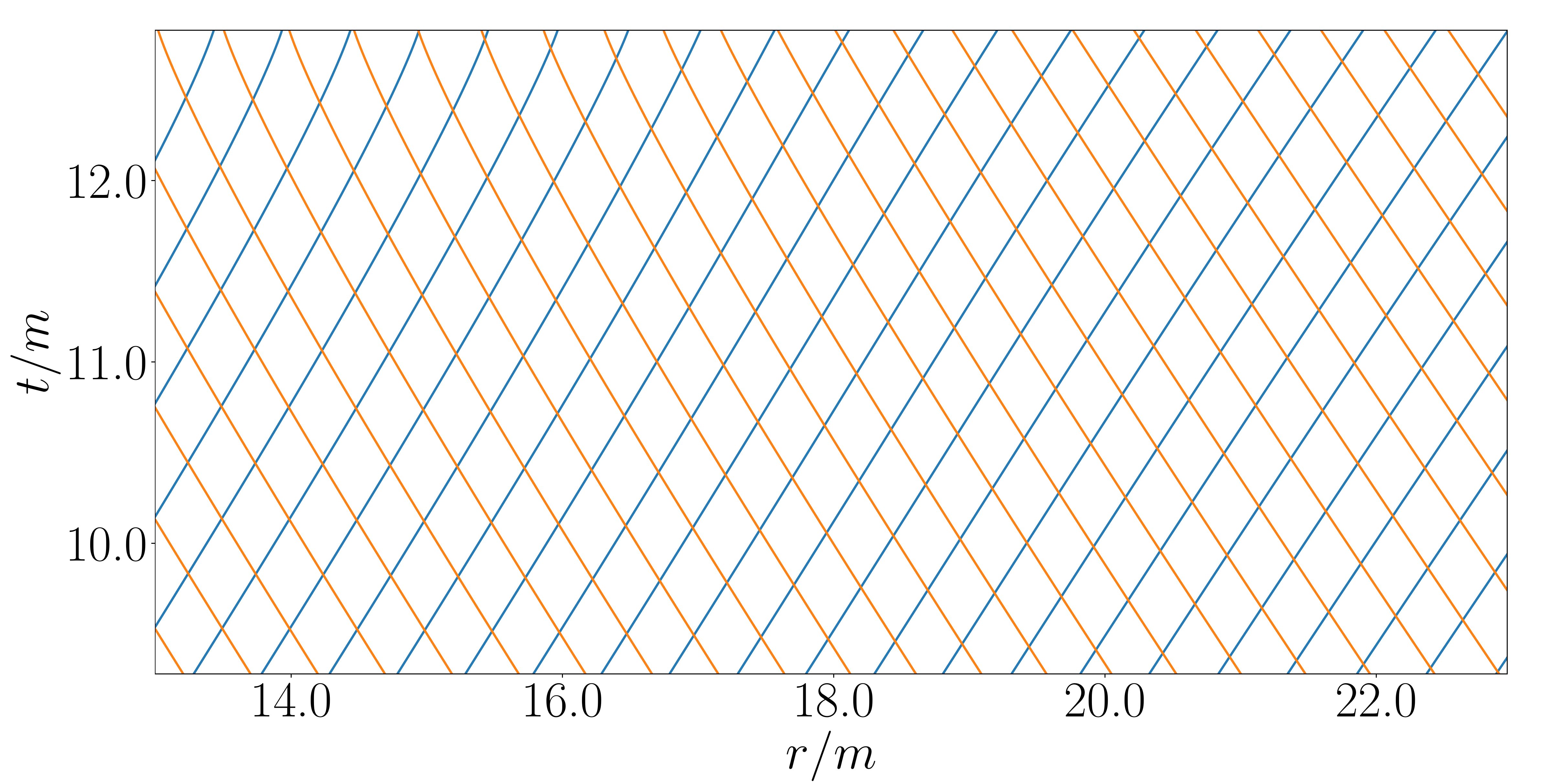}}}
	\caption{
        Characteristic and null lines from a case with identical initial data as
        in Figure \ref{fig:elliptic_forming_regions_positive_lambda}, but here
        $\lambda=-50$ (opposite sign). Qualitatively the figures
        are similar, but notice the different vertical and horizontal scales.
}
    \label{fig:elliptic_forming_regions_negative_lambda}%
\end{figure}

\begin{figure}%
	\centering
    \includegraphics[width=0.9\textwidth]{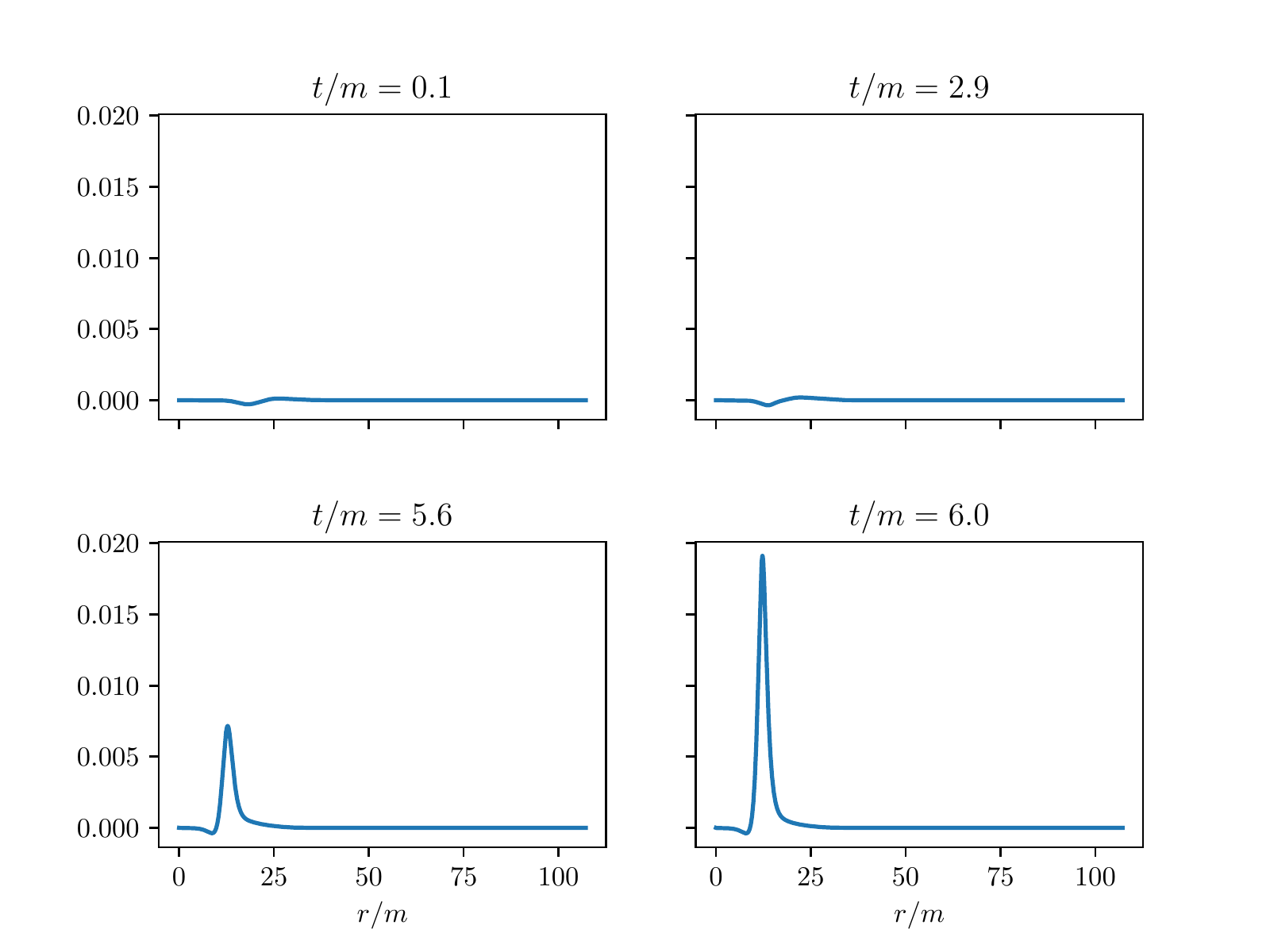}
	\caption{
        The Ricci scalar $R$ at several different times from the strong coupling,
        weak field run with $\lambda=50$ (as in Figure~\ref{fig:elliptic_forming_regions_positive_lambda}).
	The lower left panel corresponds to the time the elliptic region first forms at $r/m \sim 12.5$.
	}
    \label{fig:ricci_panels}%
\end{figure}

\begin{figure}
	\centering
    \subfloat[$|R|_{\infty}$ at three resolutions]{{\includegraphics[width=0.6\textwidth]{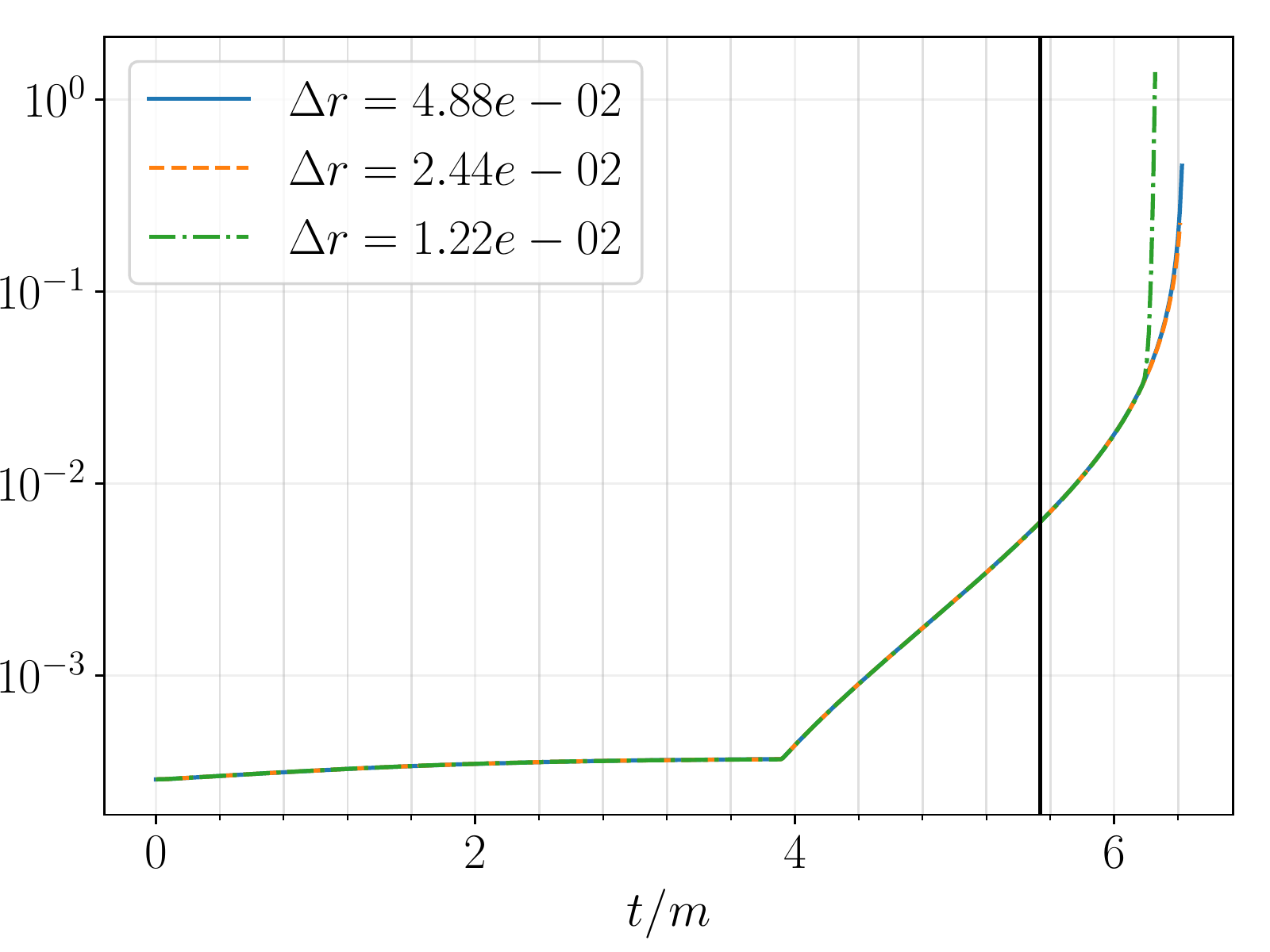} }}
   \\ 
    \subfloat[Convergence of $R$]{{\includegraphics[width=0.6\textwidth]{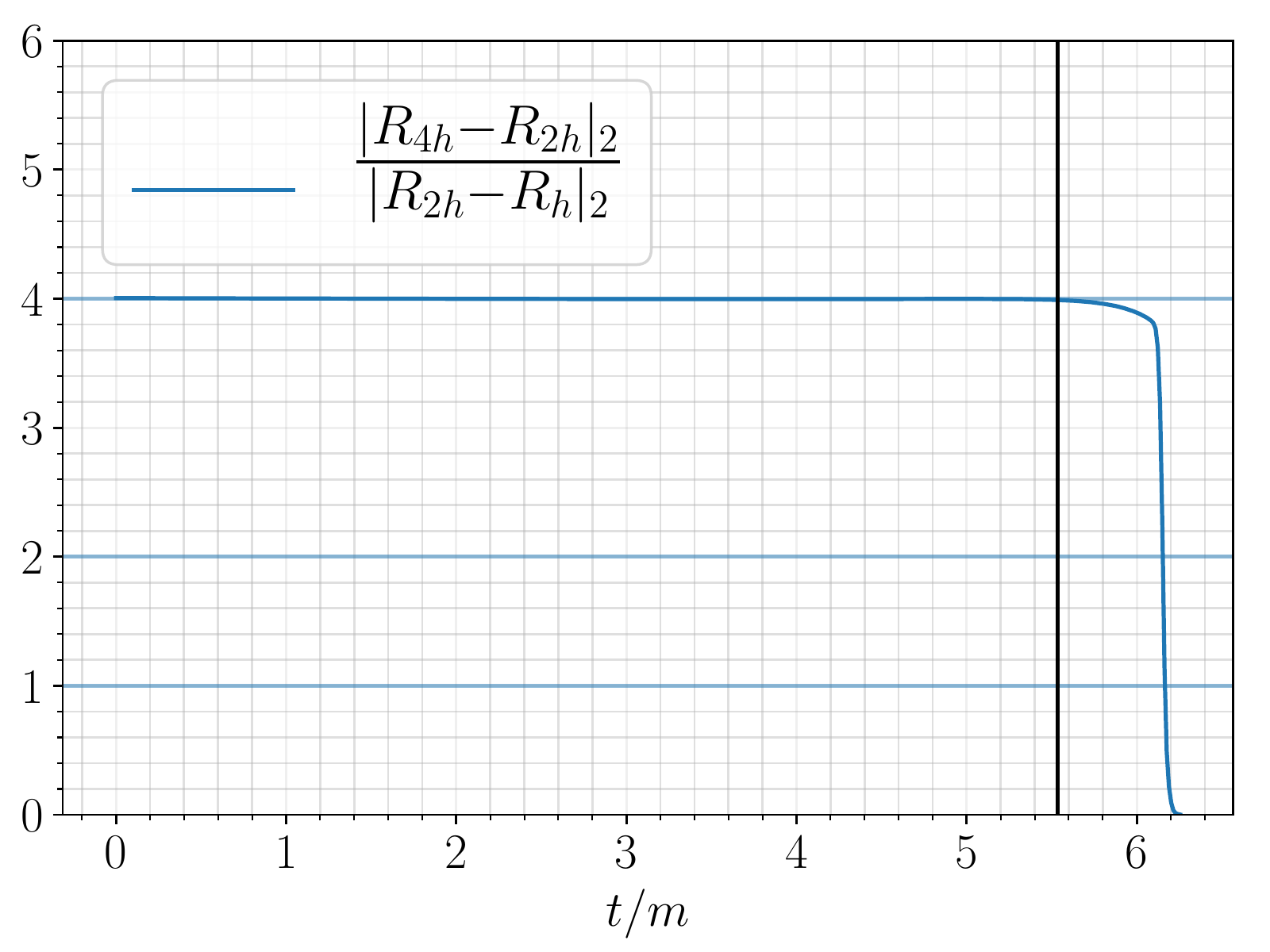} }}
    \\
	\caption{
        The top panel shows the $L_\infty$ norm of the Ricci scalar $R$ with time
        from the strong coupling, weak field $\lambda=50$ case (as depicted in Figure \ref{fig:ricci_panels} above).
        To demonstrate convergence, data from 3 different resolutions are shown.
        The bottom panel
		shows a corresponding convergence factor (computed with the $L_2$ norm),
        consistent with second order convergence prior to formation
        of the elliptic region (denoted by the vertical line at $t/m\sim 5.6$).
        This shows we are converging to a finite value of $R$ at the time the sonic
        line is first encountered. Since following this time the EdGB ($P,Q$) subsystem
        becomes ill-posed treated as a hyperbolic PDE system, as indicated by
        the drop in the convergence factor (which in theory will happen more rapidly
        with ever increasing resolution), we cannot say anything conclusive about
        some putative analytic solution at any given resolution beyond this. 
	}
    \label{fig:ricci_infinity}%
\end{figure}

	For both cases (i.e. independent of the sign of $\lambda$), 
the solutions develop an elliptic region---see Figures \ref{fig:elliptic_forming_regions_positive_lambda} and
\ref{fig:elliptic_forming_regions_negative_lambda}.
Interestingly, even though the sign of the Gauss-Bonnet coupling $\lambda$
has little effect on the ADM mass of the spacetime
it significantly affects when and where the elliptic region forms, as is evident
in these figures.
Preceding formation of this elliptic region,
the outgoing scalar field characteristic speeds near it become negative,
akin to trapped surface formation in GR gravitational collapse. However,
the spacetime outgoing null characteristic speeds $e^{A-B}$ remain positive
and well away from zero throughout the integration domain.
Hence, this elliptic region
is not ``censored'' by spacetime causal structure (the ADM
mass of the spacetimes are below the smallest known static
black hole solutions in EdGB gravity~\cite{Kanti:1995vq,Sotiriou:2013qea,Sotiriou:2014pfa},
and even so, the elliptic regions form well outside $r=2m$, so it does not
seem plausible that some spacetime trapped region could eventually form to hide the
elliptic region from asymptotic view).
At the sonic line bounding the hyperbolic from elliptic region, 
all field variables are smooth and finite. In particular, there is no geometric or scalar field
singularity that might otherwise have suggested the classical theory has already ceased to
give sensible predictions prior to this; see Figures \ref{fig:ricci_panels} and \ref{fig:ricci_infinity}
that show the Ricci scalar as an example.


That the character of the $(P,Q)$ subsystem is hyperbolic in some regions
of the spacetime, and elliptic in others (separated by the parabolic sonic line),
means the EdGB equations can be of {\em mixed type} (note of course that this is different
from coupled elliptic/hyperbolic systems often encountered in GR evolution, 
where some equations are elliptic, others hyperbolic, but each equation
maintains its definite character throughout the domain).
Mixed-type equations are not as common in the literature, but do arise
in several situations, such as steady transonic flow (see for example~\cite{otway2015elliptic},
which also discusses other areas where mixed type equations appear). 
There are two canonical mixed type equations that at least locally (near the sonic line)
are expected to capture the nature of most mixed type equations : the
Tricomi equation
\begin{equation}
\partial_y^2u(x,y) + y\ \partial_x^2u(x,y)=0, 
\end{equation}
and the Keldysh equation 
\begin{equation}
\partial_y^2u(x,y) + \frac{1}{y} \partial_x^2u(x,y) =0.
\end{equation}
These equations are hyperbolic/parabolic/elliptic for $y<0\ $/$\ y=0\ $/$\ y>0$.
The main qualitative differences between these two equations
are how the characteristics in the hyperbolic region meet the 
parabolic sonic line, and how the characteristic speeds
become imaginary. For the Tricomi equation, the characteristics intersect
the sonic line orthogonally, with the corresponding speeds going imaginary
passing though zero there. 
For the Keldysh equation, the characteristics intersect
the sonic line tangentially, with the characteristic speeds
diverging there before becoming imaginary.
This affects the degree of smoothness one can generally
expect for solutions to these equations, with the Keldysh equation 
having weaker regularity of solutions on the sonic line (see e.g. ~\cite{otway2015elliptic}).
Though the EdGB equations are vastly more complicated than
these simple prototypes, at least based on the way the characteristics
intersect the sonic line, as is apparent in Figures \ref{fig:elliptic_forming_regions_positive_lambda} and
\ref{fig:elliptic_forming_regions_negative_lambda},
and that the characteristic
speeds go to zero there, it appears that the
EdGB equations are of Tricomi type. This is typical for all
cases we have run where an elliptic region forms 
(though interestingly, for a certain class of $P(X)$ Horndeski
theories in similar collapse scenarios, ~\cite{Bernard:2019fjb} find
either Tricomi or Keldysh behavior approaching the sonic line,
depending upon the initial data in the hyperbolic region).

That the mixed type behavior here appears to be of Tricomi type is a somewhat promising
sign for EdGB gravity in terms of regularity on the sonic line (as
we explicitly find in our solutions); however, that an elliptic
region forms regardless of its type is problematic for the theory
being capable of serving as a viable, physical model that can make predictions in the
sense of possessing a well-posed initial value problem (for further discussion
on this see~\cite{PhysRevD.99.084014}).

\subsection{\label{sec:MisnerSharpMass_and NCC}
	Misner-Sharp mass and the null convergence condition}
\begin{figure}%
    \centering
    \includegraphics[width=0.8\textwidth]{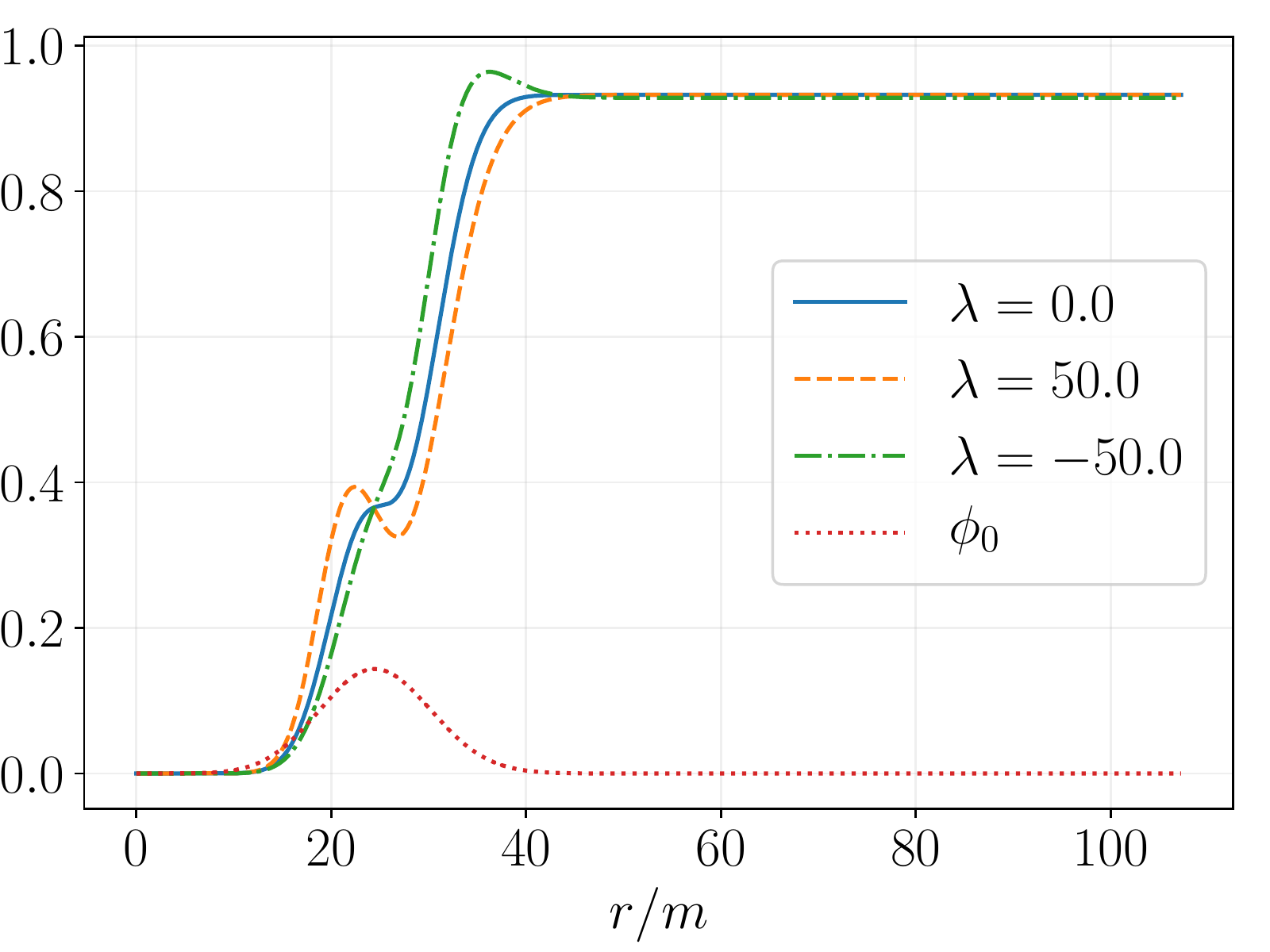} 
	\caption{
	The initial Misner-Sharp mass $m_{MS}$ (\ref{eq:misner_sharp_definition}) and
        scalar field (\ref{eq:initial_data_family_1}) profile
        for the strong coupling, weak field cases with $\lambda=\pm 50$ 
        (initial data as in Figures \ref{fig:elliptic_forming_regions_positive_lambda} and \ref{fig:elliptic_forming_regions_negative_lambda}),
        together with a $\lambda=0$ case for reference. The initial scalar field data
		is the same for all three $\lambda$ runs.
        We see that $m_{MS}$ is not always monotonically increasing as in GR ($\lambda=0$),
        though interestingly despite significant variations with $\lambda$ in the interior profile
        of $m_{MS}$, the asymptotic values are largely insensitive to $\lambda$
        .}
    \label{fig:misner_sharp_masses_different_lambda}%
	\centering
\end{figure}

Figure \ref{fig:misner_sharp_masses_different_lambda} is a plot
of the initial Misner-Sharp mass profiles for the two strong coupling ($\lambda=\pm 50$) weak field
cases, together with initial data with equivalent parameters for the GR ($\lambda=0$) case.
As discussed in Section \ref{sec:quasi_local_mass}, 
we may interpret $\partial_rm_{MS}(t,r)/4\pi r^2$ as an effective local energy density at $(t,r)$.
As is apparent in the figure, for EdGB gravity there are cleary regions where this energy density
is negative (this phenomenon has been notice before in static solutions, see e.g. \cite{Kanti:1995vq}).
Despite large variations in $m_{MS}$ in the interior as the Gauss-Bonnet coupling $\lambda$ is varied, we find
that the ADM mass (estimated by evaluating the Misner-Sharp mass
at $r=r_{max}$) depends much more weakly on $\lambda$.
With fixed initial data ($a_0=0.02$, $w_0=8$, $r_0=20$), the
ADM mass changes by at most 1 part in $10^3$ as $\lambda$
varies from -75 to 75, where we estimate the numerical error in this
quantity to be less than $1$ part in $10^4$ (from truncation error and
finite radius effects).

\begin{figure}%
    \centering
    \includegraphics[width=0.9\textwidth]{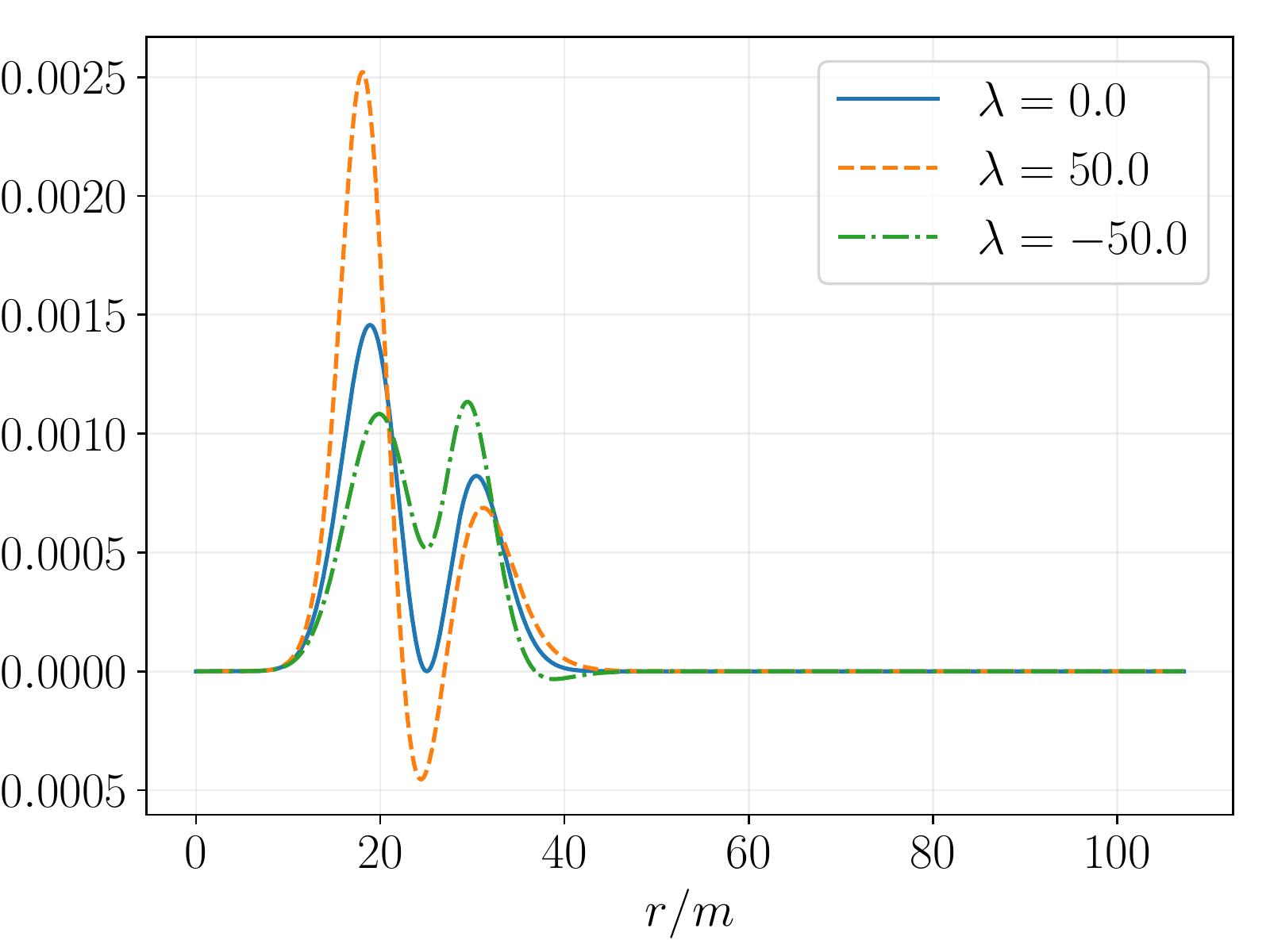}
	\caption{
        The NCC (\ref{eq:null_convergence_condition}) evaluated at $t=0$ 
        for the two strong coupling, weak field cases 
        (as in Figures \ref{fig:elliptic_forming_regions_positive_lambda} and \ref{fig:elliptic_forming_regions_negative_lambda}),
        together with the GR case ($\lambda=0$) for reference.
	The regions of NCC violation (for $\lambda\neq0$) roughly correspond to the
	regions of negative effective energy density; compare with
	Figure \ref{fig:misner_sharp_masses_different_lambda}. 
	}
    \label{fig:NCC_different_lambda}%
\end{figure}
\begin{figure}%
	\centering
    \subfloat[NCC at $t=0$ at three resolutions]{{\includegraphics[width=0.74\textwidth]{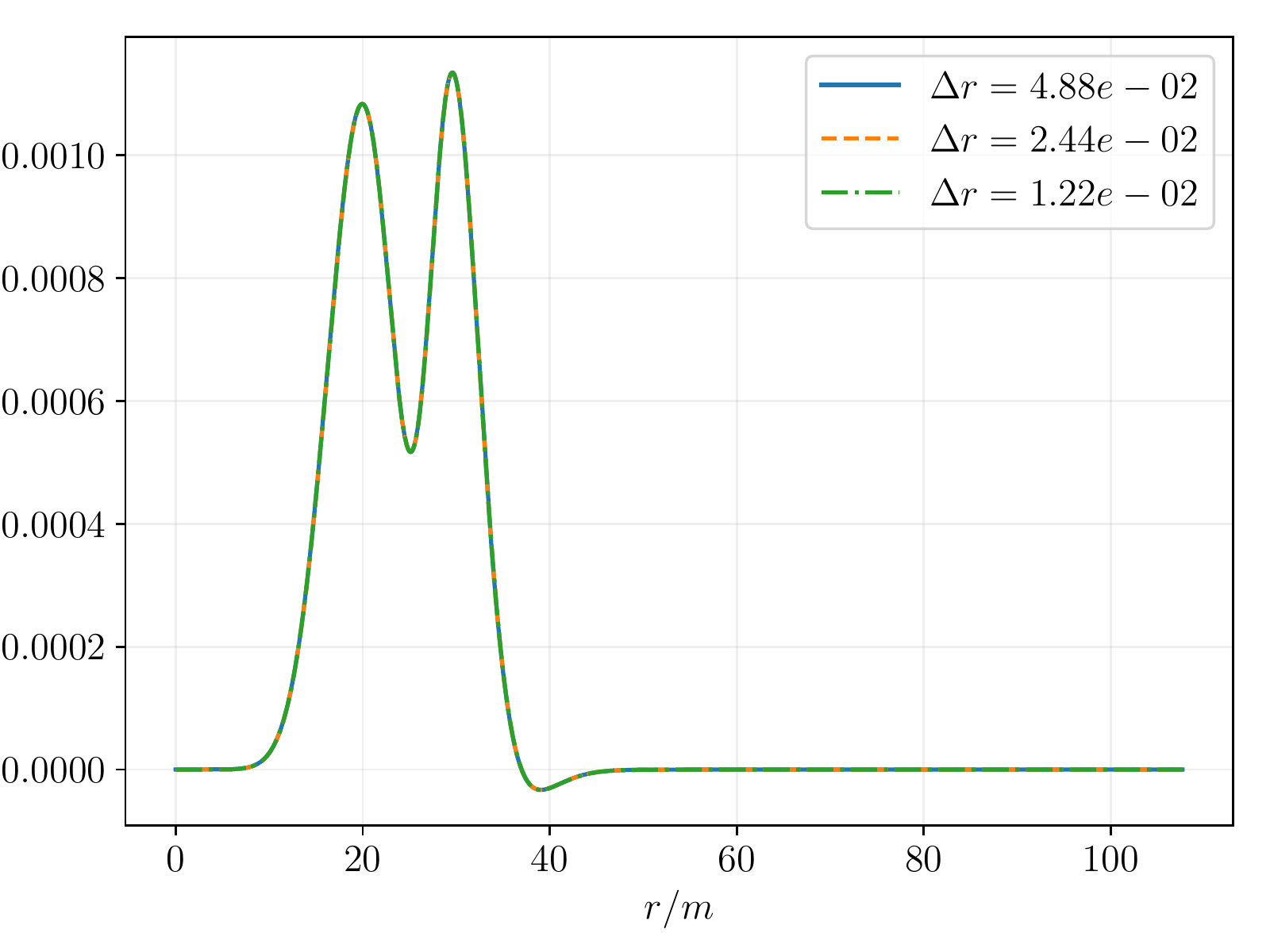} }}
    \\
    \subfloat[Convergence factor of NCC] {{\includegraphics[width=0.7\textwidth]{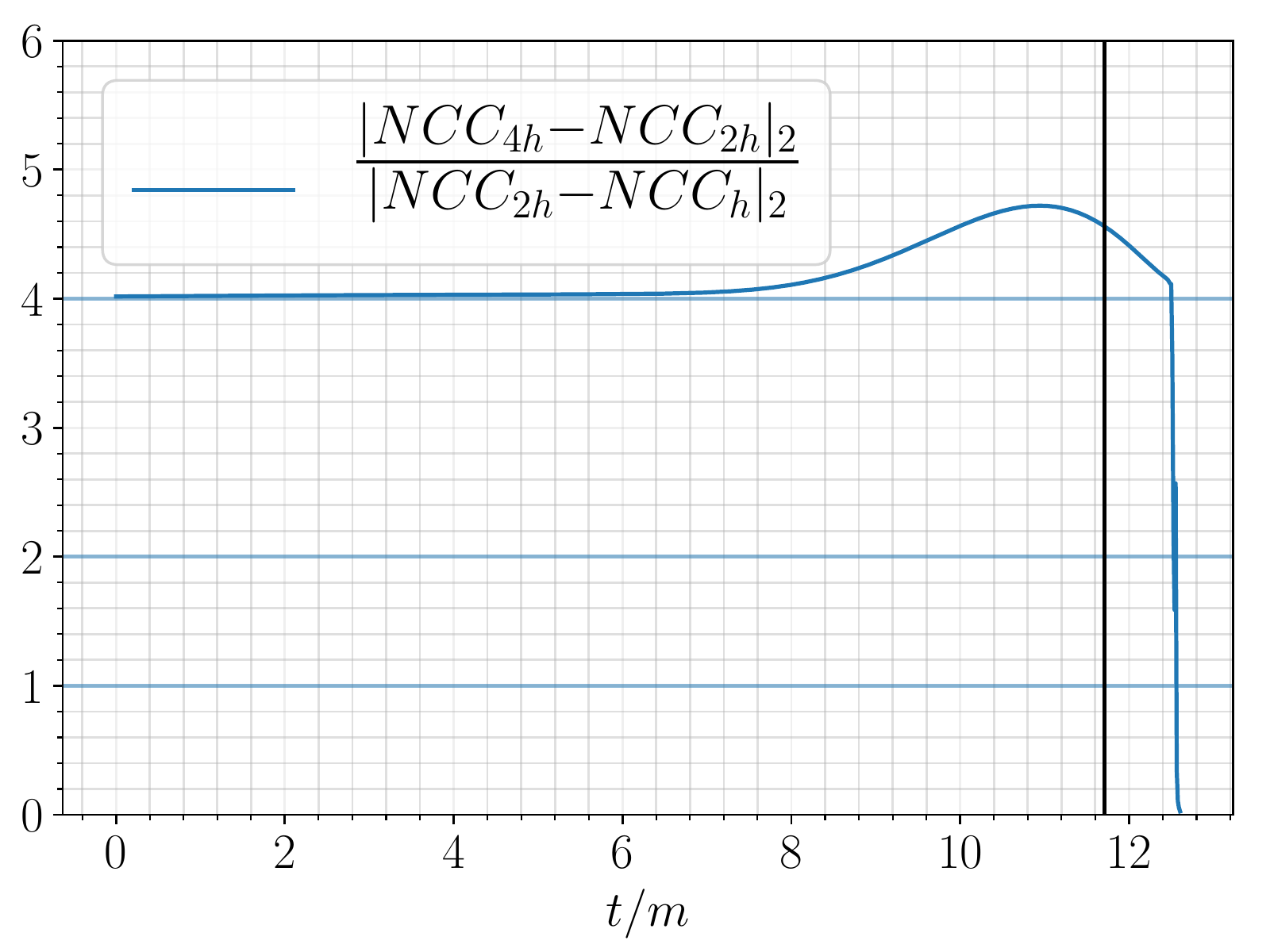} }}.
	\caption{ 
        The NCC (\ref{eq:null_convergence_condition}) 
        at $t=0$ (top) for the $\lambda=-50$ case (as in Figure \ref{fig:elliptic_forming_regions_negative_lambda})
        computed with 3 resolutions, and a corresponding convergence factor vs time (bottom), consistent
        with second order convergence of the solution. The sonic line for this case is first encountered
        at $t/m\sim 11.7$, indicated by the vertical solid line on the right panel.
}
    \label{fig:NCC_convergence_negative_lambda}%
    \centering
\end{figure}

Related to the negative effective energy densities, we find that the NCC 
(\ref{eq:null_convergence_condition}) is violated around these
regions for the non-zero $\lambda$ cases : see Figures \ref{fig:NCC_different_lambda}
and \ref{fig:NCC_convergence_negative_lambda}. We note that we find
no correlation between the existence of negative energy density regions or
regions of NCC violation and the formation of elliptic regions. While we always
observe negative energy density regions and regions of NCC violation at 
the formation of an elliptic region, we also observe those regions in simulations
where the evolution remains hyperbolic.  

\subsection{Convergence of simulations\label{sec:convergence}}
\begin{figure}%
	\centering
    \subfloat[$|E_{\vartheta\vartheta}|_2$. $\lambda=50$]{{\includegraphics[width=0.7\textwidth]{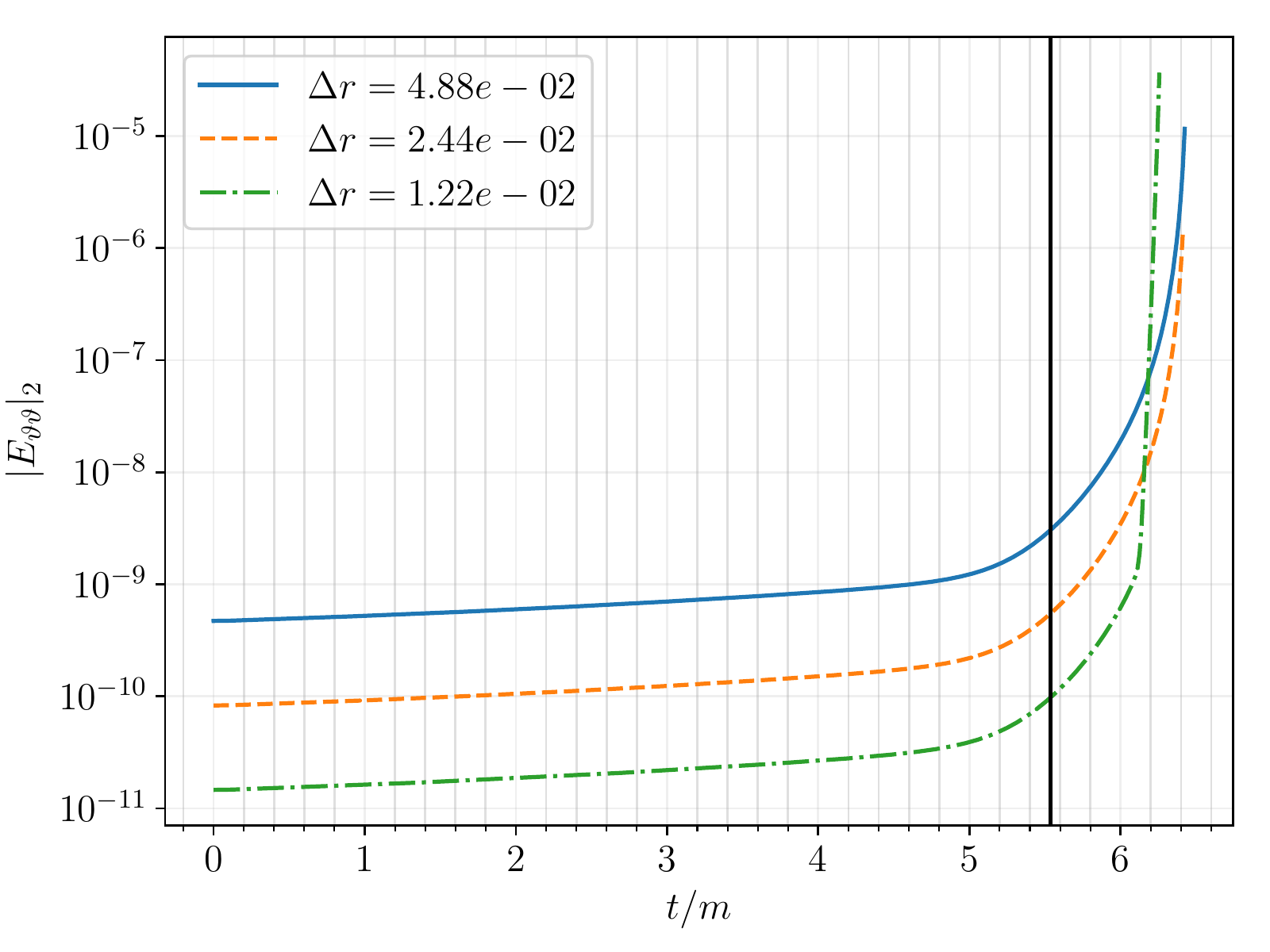} }}
    \\
    \subfloat[$|E_{\vartheta\vartheta}|_2$. $\lambda=-50$]{{\includegraphics[width=0.7\textwidth]{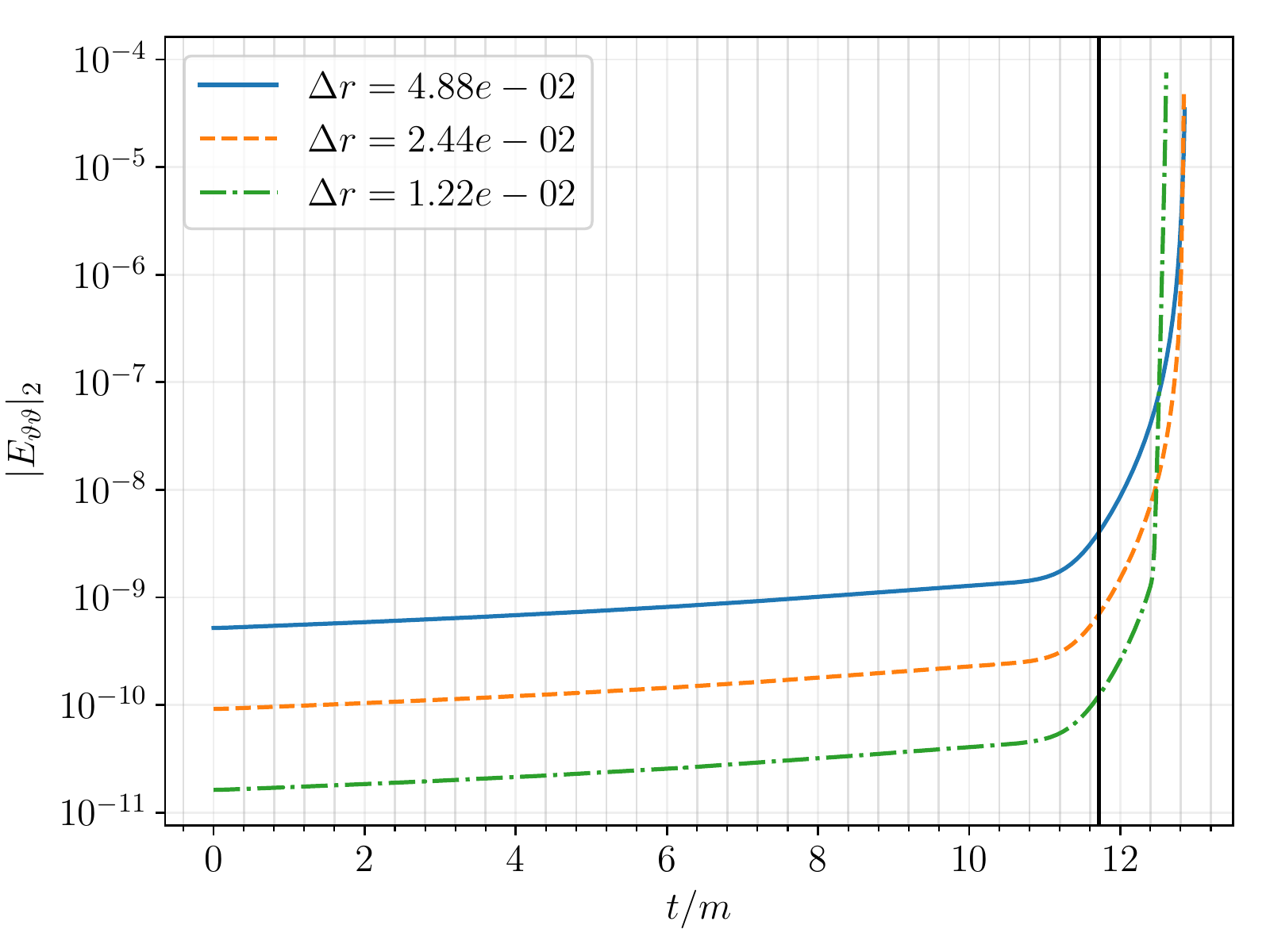} }}
	\caption{
        The $L_2$ norm of the residual of $E_{\vartheta\vartheta}$ (\ref{eq:EdGB_tensor_eom})
        for the weak field, strong coupling cases (as in Figures \ref{fig:elliptic_forming_regions_positive_lambda}
        and \ref{fig:elliptic_forming_regions_negative_lambda}). The convergence
        to zero prior to formation of the elliptic region is consistent with second
        order convergence; the growth of the residual and failure of convergence
        past this time is consistent with trying to solve a mixed type equation
        using a hyperbolic solution scheme.
}
    \label{fig:indep_res_ThTh_pos_neg_lambda}%
    \centering
\end{figure}

	In addition to convergence data we have already shown
in Figures \ref{fig:comparison_scalar_field_to_analytic_value},\ref{fig:ricci_infinity} and 
\ref{fig:NCC_convergence_negative_lambda}, in Figure \ref{fig:indep_res_ThTh_pos_neg_lambda}
we show convergence plots from the two strong coupling ($\lambda=\pm 50$)
cases for the independent residual of the $\vartheta\vartheta$ component
of the EdGB equations (\ref{eq:EdGB_tensor_eom}). That this converges
to zero (at second order) prior to formation of an elliptic region is a rather non-trivial
check of the correctness of our solution, as $E_{\vartheta\vartheta}$ depends on
temporal and spatial gradients of all variables $(P,Q,A,B)$ in the problem
(the EdGB equations, as GR, are over-determined, allowing for such 
non-trivial checks of a solution obtained from a complete subsystem
of PDEs). That we loose convergence after formation of the elliptic
region is consistent with the fact that we are attempting to
solve a mixed type equation using hyperbolic methods, which are
not well-posed in the elliptic region (for more discussion on this see \cite{PhysRevD.99.084014}).

	We report that in addition to the convergence tests we have discussed
and presented in this paper,
we achieved second order convergence before the formation of elliptic regions
for all of the fields and diagnostics we implemented in our simulations,
including the EdGB and null characteristics (as shown for exmaple
in Figure \ref{fig:elliptic_forming_regions_positive_lambda}), and the mass aspect,
(Figure \ref{fig:misner_sharp_masses_different_lambda}). Interestingly,
as with the regions of NCC violation, with the resolutions reported in this
paper we resolve the regions of \emph{negative} energy density ($\partial_rm_{MS}<0$)
seen in Figure \ref{fig:misner_sharp_masses_different_lambda}. 

\section{\label{sec:conclusion}Conclusion}
	In this paper we presented studies of numerical solutions of EdGB gravity 
in spherical symmetry in gravitational collapse-like scenarios, focusing on how 
properties of the solutions differ from similar situations in Einstein gravity minimally
coupled to a massless scalar field. For sufficiently weak EdGB coupling we
find results similar to GR : a weak field limit where the scalar field
pulse disperses beyond the integration domain, and a strong field
were a geometric horizon begins to form. In the latter scenarios,
the EdGB scalar begins to grow outsize the nascent horizon in a manner consistent with known static ``hairy'' 
BH solutions. In the strong EdGB coupling regime, we find markedly different
behavior from GR : 
(1) the equations of motion can be of mixed type, where an initially
hyperbolic system shows development of a parabolic sonic line in a localized
region of the domain, beyond which the character of the PDEs switches
to elliptic (2) there are regions of negative effective energy density, and 
(3) there are regions where the NCC is
violated. In the cases we have studied these three properties
occur together within roughly the same region of spacetime. 
While the potential physical consequences for negative energy
density and NCC violation have been extensively discussed in the modified
gravity literature, the physical interpretation of mixed type equations 
remains largely unexplored. At the very least, mixed type behavior
signals loss of predictability in the theory in the sense of it ceasing
to possess a well-posed IVP. 

One of our main motivations for studying EdGB gravity
is to discover a viable, interesting modified gravity theory to
confront with LIGO/Virgo binary BH merger data, in particular
the part of the signals attributable to coalescence. In that regard,
our results reported here and in a companion paper~\cite{PhysRevD.99.084014} 
do not yet rule out a coupling parameter that gives a smallest possible
static BH solution of around a few solar masses (which would give the
most significant differences from GR for stellar mass BH mergers),
{\em if} we assume there is no cosmological background for the EdGB scalar 
(i.e. it is only present as sourced by curvature produced by other
matter/BHs in the universe, though even then we need to ignore
problems that might arise in the very early, pre-Big-Bang-Nucleosynthesis universe).
So we have a tentative green light to continue this line of exploration
of EdGB gravity. The next step is to solve the EdGB equations 
in spherical symmetry in a horizon penetrating coordinate system. 
This will allow us to begin addressing issues of long term, non-linear
stability of hairy BHs, and perform a more thorough investigation
of the strong field, strong coupling regime. Considering the qualitatively
different behavior for GR we see in the EdGB simulations in the strong
coupling regime, it would also be interesting to understand
the nature of critical collapse in EdGB gravity, where (at least
in GR) one can dynamically evolve from smooth initial data to regions
of potentially unbounded curvature. We are presently
working on a code to study this phenomena as well.

\ack

We thank J. Anderson, L. Lehner, V. Paschalidis, I. Rodnianski, L. Stein, 
and M. Taylor for useful conversations on aspects of 
this project. We thank the organizers of the workshop
`Numerical Relativity beyond
General Relativity' and
the Centro de Ciencias de Benasque Pedro Pascual, 2018,
where we completed some of the work presented here.
F.P. acknowledges support from NSF grant PHY-1607449, the Simons
Foundation, and the Canadian Institute For Advanced Research (CIFAR).
Computational resources were provided by the
Feynman cluster at Princeton University.

\appendix

\section{\label{sec:EdGB_eom}Derivation of dilaton Gauss-Bonnet tensor}
	Here we derive equations of motion for the dilaton Gauss-Bonnet
term  
\begin{eqnarray}
	S_{GB} = 
	\int d^4x\sqrt{-g} 
	 f(\phi) \mathcal{G} 
	.
\end{eqnarray}
	Varying the Gauss-Bonnet term with respect to the metric, we have
\begin{eqnarray}
\fl	\delta\left(
	\sqrt{-g}\frac{1}{4}
	\delta^{\rho\kappa\alpha\beta}_{\lambda\sigma\gamma\delta}
	R^{\lambda\sigma}{}_{\rho\kappa}R^{\gamma\delta}{}_{\alpha\beta}
	\right)
	=
	\nonumber \\
	\sqrt{-g}\frac{1}{4}
	\delta^{\rho\kappa\alpha\beta}_{\lambda\sigma\gamma\delta}
	\left(
	2R^{\lambda\sigma}{}_{\rho\kappa}\delta R^{\gamma\delta}{}_{\alpha\beta}
	-\frac{1}{2}	
	R^{\lambda\sigma}{}_{\rho\kappa}R^{\gamma\delta}{}_{\alpha\beta}
	g_{\mu\nu}\delta g^{\mu\nu}
	\right)
	. 
\end{eqnarray}
	We focus on the variation of the Riemann tensor term:
\begin{eqnarray}
	\delta^{\rho\kappa\alpha\beta}_{\lambda\sigma\gamma\delta}
	R^{\lambda\sigma}{}_{\rho\kappa}\delta R^{\gamma\delta}{}_{\alpha\beta}
	= & 
	\delta^{\rho\kappa\alpha\beta}_{\lambda\sigma\gamma\delta}
	\left(
	R^{\lambda\sigma}{}_{\rho\kappa}R^{\gamma}{}_{\omega\alpha\beta}
	\delta g^{\omega\delta}
	+ 
	R^{\lambda\sigma}{}_{\rho\kappa}
	g^{\omega\delta}\delta R^{\gamma}{}_{\omega\alpha\beta}
	\right)
	.
\end{eqnarray}
	In four dimensions, a five index antisymmetric tensor is zero, so
we may write (c.f. Appendix A and B of \cite{tHooft:1974toh}) 
\begin{eqnarray}
\fl	\delta^{\rho\kappa\alpha\beta}_{\lambda\sigma\gamma\delta}
	g_{\omega\iota}
	R^{\lambda\sigma}{}_{\rho\kappa}R^{\gamma\iota}{}_{\alpha\beta}
	\delta g^{\omega\delta}
	=  
	\nonumber \\
	\left(
	\delta^{\rho\kappa\alpha\beta}_{\iota\sigma\gamma\delta}
	g_{\omega\lambda}
	+ \delta^{\rho\kappa\alpha\beta}_{\lambda\iota\gamma\delta}
	g_{\omega\sigma}
	+ \delta^{\rho\kappa\alpha\beta}_{\lambda\sigma\iota\delta}
	g_{\omega\gamma}
	+ \delta^{\rho\kappa\alpha\beta}_{\lambda\sigma\gamma\iota}
	g_{\omega\delta}
	\right)
	R^{\lambda\sigma}{}_{\rho\kappa}R^{\gamma\iota}{}_{\alpha\beta}
	\delta g^{\omega\delta}
	,
\end{eqnarray} 
	which implies
\begin{eqnarray}
	\delta^{\rho\kappa\alpha\beta}_{\lambda\sigma\gamma\delta}
	R^{\lambda\sigma}{}_{\rho\kappa}R^{\gamma}{}_{\omega\alpha\beta}
	\delta g^{\omega\delta}
	= &
	\frac{1}{4} 
	\delta^{\rho\kappa\alpha\beta}_{\lambda\sigma\gamma\delta}
	R^{\lambda\sigma}{}_{\rho\kappa}R^{\gamma\delta}{}_{\alpha\beta} 
	g_{\mu\nu}\delta g^{\mu\nu}
	.
\end{eqnarray}
	We conclude that in four spacetime dimensions, 
the variation of the Gauss-Bonnet term with respect to the metric is
\begin{eqnarray}
\fl	\delta\left(
	\sqrt{-g}\frac{1}{4}
	\delta^{\rho\kappa\alpha\beta}_{\lambda\sigma\gamma\delta}
	R^{\lambda\sigma}{}_{\rho\kappa}R^{\gamma\delta}{}_{\alpha\beta}
	\right)
	& =
	\sqrt{-g}\frac{1}{2}
	\delta^{\rho\kappa\alpha\beta}_{\lambda\sigma\gamma\delta}
	R^{\lambda\sigma}{}_{\rho\kappa}g^{\omega\delta}
	\delta R^{\gamma}{}_{\omega\alpha\beta}
	\nonumber \\
\fl	& =
	\sqrt{-g}
	\delta^{\rho\kappa\alpha\beta}_{\lambda\sigma\gamma\delta}
	R^{\lambda\sigma}{}_{\rho\kappa}
	g^{\omega\delta}g^{\gamma\iota}
	\nabla_{\alpha}\nabla_{\omega}\delta g_{\iota\beta}
	.
\end{eqnarray}	
	Relabeling indices, and noting that from the Bianchi identities
$
	\delta^{\rho\kappa\alpha\beta}_{\lambda\sigma\gamma\delta}
	R^{\lambda\sigma}{}_{\rho\kappa}
$ is divergenceless on all its indices
(e.g. \cite{misner1973gravitation} \S13.5) 
the variation of the dilaton Gauss-Bonnet term is
\begin{equation}
	\delta S_{GB}
	=
	-\delta^{\gamma\delta\kappa\lambda}_{\beta\alpha\rho\sigma}
	R^{\rho\sigma}{}_{\kappa\lambda}
	\left(\nabla_{\gamma}\nabla^{\alpha}f(\phi)\right)
	\delta^{\beta}_{\mu}g_{\nu\delta}\delta g^{\mu\nu}
	,
\end{equation}
	plus surface terms. Using similar manipulations as presented above,
we note that taking the divergence of
the Gauss-Bonnet tensor is
\begin{eqnarray}
\fl	\nabla^{\mu}\left(
	\delta^{\gamma\delta\kappa\lambda}_{\alpha\beta\rho\sigma}
	R^{\rho\sigma}{}_{\kappa\lambda}
	\left(\nabla_{\gamma}\nabla^{\alpha}f(\phi)\right)
	\delta^{\beta}_{\mu}g_{\nu\delta}
	\right)
	 & =
	\frac{1}{2}g_{\nu\delta}
	R^{\rho\sigma}{}_{\kappa\lambda}R_{\gamma\omega}{}^{\beta\alpha}
	\delta^{\gamma\delta\kappa\lambda}_{\alpha\beta\rho\sigma}
	\nabla^{\omega}\phi
	\nonumber \\
\fl	 & =
	- \frac{1}{2}\mathcal{G} \nabla_{\nu}f(\phi)
	.
\end{eqnarray}
so that assuming $\nabla_{\nu}\phi\neq0$, taking the divergence
of \eref{eq:EdGB_tensor_eom} gives us
\eref{eq:EdGB_scalar_eom} (the `generalized Bianchi identity' 
\cite{Tanahashi:2017kgn}). 

\section{\label{sec:EdGB_equations_of_motion}EdGB equations of motion}
	In the coordinates (\ref{eq:line_element_polar_coordinates}),
the nontrivial components of the EdGB equations of motion
\eref{eq:EdGB_tensor_eom} are 
\begin{eqnarray}
	\label{eq:EdGB_EOM:TT}
\fl	E^{(g)}_{tt} 
	\propto & 
	\left(1 + 4\lambda\left(1 - 3e^{-2B}\right)\frac{Q}{r}\right)\partial_rB
	+ \frac{e^{2B}-1}{2r}
	- \frac{1}{2}r\left(Q^2+P^2\right)
	\nonumber \\
\fl	& 
	+ 4\lambda\frac{-1+e^{-2B}}{r}
	\left(
	\partial_rQ 
	+ e^{-A-B} P \partial_tB
	\right) = 0
	, \\
	\label{eq:EdGB_EOM:TR}
\fl	E^{(g)}_{tr} 
	\propto & 
	\left(1 + 4\lambda\left(1-3e^{-2B}\right)\frac{Q}{r}\right)\partial_tB
	- \frac{1}{2}r e^{A-B} QP
	\nonumber \\
\fl	& + 
	4\lambda e^{A-B}\frac{1-e^{-2B}}{r}
	\left(
	P\partial_rB - \partial_rP
	\right) = 0
	, \\
	\label{eq:EdGB_EOM:RR}
\fl	E^{(g)}_{rr}
	\propto &
	\left(1 + 4\lambda\left(1-3e^{-2B}\right)\frac{Q}{r}\right)\partial_rA
	+ \frac{1-e^{2B}}{2r} - \frac{1}{2}r\left(Q^2+P^2\right)
	\nonumber \\
\fl	& + 
	4\lambda e^{-A-B}\frac{e^{2B}-1}{r}
	\left(
	P \partial_tB - \partial_tP
	\right) = 0
	, \\
	\label{eq:EdGB_EOM:ThTh}
\fl	E^{(g)}_{\vartheta\vartheta}
	\propto & 
	\left(-1 + 8\lambda e^{-2B}\frac{Q}{r}\right)
	\left(
	\partial_t^2B
	- e^{2A-2B}\partial_r^2A
	+ e^{2A-2B}\left(\partial_rA\right)^2
	+ \partial_tA \partial_tB
	\right)
	\nonumber \\
\fl	& - 
	\left(1 + 8\lambda e^{-2B}\frac{Q}{r}\right)\left(\partial_tB\right)^2
	+ 8\lambda e^{A-3B}P
	\left(
	\frac{\partial_rA}{r} - \frac{\partial_rB}{r} + 2 \frac{\partial_rP}{r}
	\right)
	\partial_tB
	\nonumber \\
\fl	& + 
	e^{2A-2B}
	\left(
	\frac{1-e^{-4B}}{r} + 24\lambda e^{-2B}\frac{Q}{r}\partial_rB 
	\right)
	\partial_rA
	+ e^{2A-2B}\frac{\partial_rB}{r}
	\nonumber \\
\fl	& 
	+ 8\lambda e^{2A-4B}\frac{\partial_rQ\partial_rA}{r}
	+ 8\lambda e^{A-3B}\frac{\partial_rB\partial_tP}{r}
	+ \frac{1}{2}e^{2A-2B}\left(Q^2-P^2\right) = 0
	.
\end{eqnarray}
	In the language of the $3+1$ ADM formalism
(\ref{eq:EdGB_EOM:TT}) is the Hamiltonian constraint,
(\ref{eq:EdGB_EOM:TR}) is the momentum constraint, and
(\ref{eq:EdGB_EOM:RR}) and (\ref{eq:EdGB_EOM:ThTh}) are part
of the evolution equations for the extrinsic curvature of spacelike
slices with normal vector $n^{\mu}=\left(e^{-A},0,0,0\right)$.

	\Eref{eq:EdGB_tensor_eom} for the EdGB scalar is
\begin{eqnarray}
\label{eq:EdGB_EOM:P}
\fl	E^{(P,\phi)} 
	\equiv &
	\partial_tP - \frac{1}{r^2}\left(r^2e^{A-B}\partial_rQ\right)
	- 8\lambda e^{-A-B}\frac{1+e^{2B}}{r^2}\left(\partial_tB\right)^2
	+ 8\lambda e^{A-3B}\frac{3-e^{2B}}{r^2}\partial_rA\partial_rB
	\nonumber \\
\fl	& 
	+ 8 \lambda e^{-A-B}\frac{1-e^{2B}}{r^2}
	\left(
		\partial_r^2B - \partial_tA\partial_tB - e^{-2B}\partial_rA
	-	e^{-2B}\left(\partial_rA\right)^2
	\right)
	= 0
	,
\end{eqnarray}
	and the evolution equation for the constraint $\partial_r\phi=Q$ is
\begin{equation}
\label{eq:EdGB_EOM:Q}
	E^{(Q)} 
	\equiv
	\partial_tQ
	- \partial_r\left(e^{A-B}\partial_rP\right)
	= 0
	.
\end{equation}
	When $\lambda=0$, it is clear from
\eref{eq:EdGB_EOM:TT} and \eref{eq:EdGB_EOM:TR} that the gravity degrees of
freedom, $A$ and $B$, are fully constrained. All the dynamics are driven
by \eref{eq:EdGB_EOM:P}. The addition of the EdGB
tensor terms introduces $\partial_tP$ and $\partial_tB$ terms into the
constraint equations. The Gauss-Bonnet scalar introduces
second derivative terms as well as $\partial_tB$, $\partial_tA$ terms
to \eref{eq:EdGB_EOM:P}. These new $\partial_tA$ and $\partial_tB$
terms appear to change the PDE character of the EdGB field equations versus
the GR field equations. As it turns out though, we can use algebraic
combinations of
\eref{eq:EdGB_EOM:ThTh} to remove second derivative and $\partial_tA$ terms,
and \eref{eq:EdGB_EOM:TR} to remove $\partial_tB$ terms from
Equations \eref{eq:EdGB_EOM:TT}, \eref{eq:EdGB_EOM:TR}, 
and \eref{eq:EdGB_EOM:P}. Doing so leads us to Equations
\eref{eq:EdGB_solved:rDer_A}-\eref{eq:EdGB_solved:tDer_P}.
\section{\label{sec:characteristics_method_two}
	A second procedure to compute the characteristics}
	Here we present another procedure we used to calculate the
characteristics of the propagating degree of freedom for the EdGB system
in spherical symmetry. Instead of substituting in for $\partial_rA$ and
$\partial_rB$ at the level of the full equations of motion, we first compute
the full principal symbol and then substitute them in from the constraints.
We find that this method is more numerically stable
near the origin at high resolutions. This is most likely because the 
length the equations to be evaluated in each component
of the principal symbol in this method are much shorter than they are
in the other, which makes them less susceptible to floating point roundoff
errors. Both methods produce equivalent results to within truncation error.

This procedure to compute the characteristics goes as follows:
we consider the full system of equations
\eref{eq:EdGB_solved:rDer_A}-\eref{eq:EdGB_solved:tDer_P}; which
take the following form 
\begin{equation}
	E^I\left(v^J,\partial_av^K\right)
	= 0
	,
\end{equation}
	where now $I,J,K$ index the fields $(A,B,Q,P)$: the equations
$E^{(Q)}$ and $E^{(P)}$ retain the terms 
$\partial_rA$ and $\partial_rB$. The characteristic
matrix for the full system is
\begin{eqnarray}
\label{eq:symbol_method_2}
	\mathfrak{p}(\xi)
	=
	\pmatrix{%
		\mathfrak{a}\xi_t + \mathfrak{b}\xi_r &
		\mathfrak{q} \xi_r \cr
		\mathfrak{r} \xi_r &
		\mathfrak{s} \xi_r
	}
	,
\end{eqnarray}
	where
\begin{eqnarray}
	\mathfrak{a}
	\equiv &
	\pmatrix{%
		\delta E^{(Q)}/\delta\left(\partial_tQ\right) & 
		\delta E^{(Q)}/\delta\left(\partial_tP\right) \cr
		\delta E^{(P)}/\delta\left(\partial_tQ\right) &
		\delta E^{(P)}/\delta\left(\partial_tP\right) 
	}
	, \\
	\mathfrak{b}
	\equiv &
	\pmatrix{%
		\delta E^{(Q)}/\delta\left(\partial_rQ\right) & 
		\delta E^{(Q)}/\delta\left(\partial_rP\right) \cr
		\delta E^{(P)}/\delta\left(\partial_rQ\right) &
		\delta E^{(P)}/\delta\left(\partial_rP\right) 
	}
	, \\
	\mathfrak{q}
	\equiv &
	\pmatrix{%
		\delta E^{(Q)}/\delta\left(\partial_rA\right) & 
		\delta E^{(Q)}/\delta\left(\partial_rB\right) \cr
		\delta E^{(P)}/\delta\left(\partial_rA\right) &
		\delta E^{(P)}/\delta\left(\partial_rB\right) 
	}
	, \\
	\mathfrak{r}
	\equiv &
	\pmatrix{%
		\delta E^{(A)}/\delta\left(\partial_rQ\right) & 
		\delta E^{(A)}/\delta\left(\partial_rP\right) \cr
		\delta E^{(B)}/\delta\left(\partial_rQ\right) &
		\delta E^{(B)}/\delta\left(\partial_rP\right) 
	}
	, \\
	\mathfrak{s}
	\equiv &
	\pmatrix{%
		\delta E^{(A)}/\delta\left(\partial_rA\right) & 
		\delta E^{(A)}/\delta\left(\partial_rB\right) \cr
		\delta E^{(B)}/\delta\left(\partial_rA\right) &
		\delta E^{(B)}/\delta\left(\partial_rB\right) 
	}
	.
\end{eqnarray}
	Provided $\mathfrak{s}$ is invertible\footnote{Note that when
$\lambda=0$, $\mathfrak{s}$ is the identity matrix. In practice,
we have never
encountered a situation where $\mathfrak{s}$ is not invertible.},
we can use Gaussian elimination
to write the characteristic equation as
\begin{equation}
	\mathrm{Det}\left(\mathfrak{p}\right)
	=
	\mathrm{Det}\left(-\mathfrak{i}c + \mathfrak{c}\right)\xi_r^2
	,
\end{equation}
	where $c\equiv-\xi_t/\xi_r$, $\mathfrak{i}$ is the identity matrix, and
\begin{equation}
	\mathfrak{c}
	\equiv
	\mathfrak{a}^{-1}
	\cdot
	\left(
	\mathfrak{b} - \mathfrak{q}\cdot\mathfrak{s}^{-1}\cdot\mathfrak{r}
	\right)
	.
\end{equation}
	The two characteristics given by $\xi_r=0$ define the characteristic
surfaces for
the constrained degrees of freedom. The characteristics
for the dynamical degree of freedom are determined by solving the nontrivial
determinant; we then find that
the characteristic speeds for this degree of freedom are
given by the eigenvalues of $\mathfrak{c}$,
\begin{equation}
	c_{\pm}
	=
	\frac{1}{2}\left(
		\mathrm{Tr}\left(\mathfrak{c}\right)
	\pm	\sqrt{
			\mathrm{Tr}\left(\mathfrak{c}\right)^2
		-	4\mathrm{Tr}\left(\mathfrak{c}\right)
			 \mathrm{Det}\left(\mathfrak{c}\right)
		}
	\right)
	.
\end{equation}
\section{\label{sec:static_decoupled}Static decoupled EdGB solution about a 
	Schwarzschild black hole background}
	In Section \ref{sec:strong_field_weak_coupling}
we compared the profile of our scalar field to that of the
`decoupled' EdGB scalar profile about a Schwarzschild background. For 
completeness we present the calculation of the profile of $\phi$. 
In the decoupling limit
of EdGB (e.g. \cite{Benkel:2016kcq,Benkel:2016rlz})
the geometry is determined by the Einstein equations coupled 
to matter fields but not the EdGB scalar field, and the equation of motion
for the EdGB scalar is given by
\begin{equation}
\label{eq:decoupled_EdGB_scalar}
	\Box\phi + \lambda \mathcal{G} = 0
	.
\end{equation}
	We consider static solutions to this equation with 
a fixed Schwarzschild black hole background
\begin{equation}
	ds^2 
	= 
	- \left(1-\frac{2M}{r}\right)dt^2
	+ \left(1-\frac{2M}{r}\right)^{-1}dr^2
	+ r^2 \left(
		d\vartheta^2 + \mathrm{sin}^2\vartheta d\varphi^2
	\right)
	.
\end{equation}
	With this, \eref{eq:decoupled_EdGB_scalar} reduces to
\begin{equation}
	\frac{1}{r^2}
	\frac{d}{dr}\left(
		r^2\left(1-\frac{2M}{r}\right)
		\frac{d\phi}{dr}
	\right)
	+ \lambda
	\frac{48M^2}{r^6}
	= 0
	.
\end{equation}
	Imposing regularity of $\partial_r\phi$ at the geometric horizon $r=2M$,
setting $\lim_{r\to\infty}\phi=0$, and changing variables to $x\equiv r/M$, we obtain
\begin{equation}
\label{eq:decoupled_scalar_profile}
	\phi(x)
	=
	\frac{2\lambda}{M^2}
	\left(
		\frac{1}{x}
	+	\frac{1}{x^2}
	+	\frac{4}{3x^3}
	\right)
	,
\end{equation}
	which is what we compare against our numerical solutions
in Figure \ref{fig:comparison_scalar_field_to_analytic_value}.
\section{\label{sec:numerical_methods}Numerical methods}
	We implemented three different finite difference PDE solution methods to solve
equations \eref{eq:EdGB_solved:rDer_A}-\eref{eq:EdGB_solved:tDer_P},
in order gain confidence that the code crashes occurring some time after formation of sonic lines 
are due to a property of the underlying continuum equations, rather than a numerical instability 
associated with a particular discretization scheme. The first two methods, described
here, are fully constrained, the third is a partially constrained scheme, described
below in \ref{sec:partially_constrained_evo}. 
All methods we implemented treat
the ($P,Q$) subsystem as hyperbolic, and
are (globally) second order accurate with fixed time and 
spatial steps. The two hyperbolic methods for $(P,Q$) we
developed are an iterative Crank-Nicolson scheme (CN), and a 
fourth order in time Runge-Kutta (method of lines) scheme (RK4). We ran simulations with CFL numbers that varied from
$10^{-2}$ to $0.5$. The different methods all give
the same results to within truncation error, and once the elliptic region forms
all crash in a qualitatively similar manner (growth of short wavelength solution components
within the elliptic region at a rate proportional to their wave number; note though that
the since our initial data is smooth, these short wavelength components
are sourced by truncation error for the most part, and their ``initial'' amplitudes
on the sonic line therefore decrease with resolution).
This gives us confidence that the crashes
are due to trying to solve a mixed type equation using hyperbolic methods, which are
not well-posed in elliptic regions.

We use the notation
$f^n_j$ for a discretized field, where $n$ stands for the time step and $j \in 0..N_r-1$ is the index within the 
spatial grid with $N_r$ points. The basic iteration loop we use for both the CN and RK4 evolution
schemes,
solving for the unknowns at time step $n+1$ given data at time step $n$, is as follows:
\begin{enumerate}
\item
Initialize time step $n+1$ values for the fields $A$, $B$, $Q$, and $P$ with
their values at time step $n$ (this step is unnecessary for the RK4 scheme).
\item
For the CN scheme (\ref{sec:hyperbolic_solvers_iter}) 
perform one step of a Newton iteration to correct 
the unknown values of $Q^{n+1}_j,P^{n+1}_j$; for the RK4
integration (\ref{sec:hyperbolic_solvers_rk4}) take the next
substep of the RK4 scheme, saving the results
in temporary arrays, or $Q^{n+1}_j,P^{n+1}_j$ for the final
step. 
\item
Integrate the constraints for $A^{n+1}_j$ and $B^{n+1}_j$ given
the current values of $Q^{n+1}_j$ $P^{n+1}_j$ (or the
appropriate substep arrays when using RK4). Since
equation (\ref{eq:EdGB_solved:rDer_B})
for $B$ does not depend on $A$, we first integrate this for $B$ (\ref{sec:ode_solvers_B}), then 
substitute the result into (\ref{eq:EdGB_solved:rDer_A}) before
integrating it for $A$ (\ref{sec:ode_solvers_A}).
\item
Repeat steps (ii) and (iii) until (a) for the CN iterative
scheme the residuals for the full nonlinear
set of equations are below a tolerance set to be a few orders of magnitude smaller than
truncation error; (b) for RK4, we have completed all the RK substeps.
\item
Apply a Kreiss-Oliger filter
(e.g. \cite{KreissOligerDissipationStandardRef})
to the now known variables $Q^{n+1}_j$ and $P^{n+1}_j$.
\end{enumerate}

\subsection{\label{sec:hyperbolic_solvers_iter}CN Hyperbolic PDE solver for $Q$ and
$P$}
For the iterative methods we employ a Crank-Nicolson discretization in time (see e.g.\cite{gustafsson1995time}),
where the equations \eref{eq:EdGB_solved:rDer_A} and \eref{eq:EdGB_solved:rDer_A} are discretized
at a time half way between time steps $n$ and $n+1$, which we denote
as time step $n+1/2$.  
Explicitly, we replace each field $f$
and its gradients with the following stencils 
\begin{eqnarray}
	f 
	& \to 
	\frac{1}{2}\left(f^{n+1}_j + f^n_j\right)
	, \\
	\partial_tf 
	& \to
	\frac{1}{\Delta t}\left(f^{n+1}_j - f^n_j\right)
	, \\
	\partial_rf
	& \to 
	\frac{1}{4\Delta r}
	\left(
		f^{n+1}_{j+1} - f^{n+1}_{j-1} + f^n_{j+1} - f^n_{j-1}
	\right)
\end{eqnarray}

We define the residual and field vectors $\mathcal{R}_k$ and $v_{k}$ respectively via
\begin{eqnarray}
	\mathcal{R}_{2j}
	& \equiv	
	\left(E^{(Q)}\right)^{n+1/2}_j	
	, \\
	\mathcal{R}_{2j+1}
	& \equiv
	\left(E^{(P)}\right)^{n+1/2}_j	
	, \\
	v_{2j}
	& \equiv	
	Q^{n+1}_j
	, \\
	v_{2j+1}
	& \equiv	
	P^{n+1}_j
	, 
\end{eqnarray}
where $0<k<2(N_r-1)$.
	For the iteration step (ii) above we compute the linear correction $\delta v_j$ by solving
the following matrix equation
\begin{equation}
\label{eq:numerical_methods_matrix_equation_QP}
	\mathcal{J}_{ij}\delta v_j + \mathcal{R}_i 
	= 0
	,
\end{equation}
	for $\delta v_j$, where
\begin{equation}
	\mathcal{J}_{ij}
	\equiv
	\frac{\delta \mathcal{R}_i}{\delta v_j}
	.
\end{equation}
	We invert the matrix $\mathcal{J}_{ij}$ in two different ways.
For the first method we 
directly solve \eref{eq:numerical_methods_matrix_equation_QP} with a banded
matrix solver (the LAPACK routine dgbsv \cite{lapack_manual}).
For the second method we solve \eref{eq:numerical_methods_matrix_equation_QP} 
with Gauss-Seidel iteration (e.g. \cite{golub2013matrix}).

\subsection{\label{sec:hyperbolic_solvers_rk4}RK4 PDE solver for $Q$ and
$P$}
We use a standard fourth order in time Runge-Kutta algorithm (see e.g.~\cite{atkinson2011numerical}),
so will not describe it here, but note that we still only employ
a second order accurate discretization for spatial gradients;
i.e. for each field $f$ we use the stencils
\begin{eqnarray}
\label{eq:center_f}
	f
	\to &
	f^n_j
	, \\
\label{eq:center_Dr_f}
	\partial_rf
	\to &
	\frac{1}{2\Delta r}
	\left( f^n_{j+1} - f^n_{j-1}\right)
	.	
\end{eqnarray}
For this study we are able to achieve the requisite accuracy with
second order methods and reasonable computer power, though do not
use a  second order Runge-Kutta method, with the radial differences
\eref{eq:center_f} and \eref{eq:center_Dr_f}, as it is unconditionally 
unstable for the linear wave equation, as may 
be verified by a von-Neumann stability analysis.

\subsection{\label{sec:ode_solvers_B}ODE integrator for $B$}
	\Eref{eq:EdGB_solved:rDer_B} for the $B$ field takes the schematic form
\begin{equation}
\label{eq:numerical_method_rDer_B}
	c_{(B)}\partial_rB + d_{(B)} = 0
	,
\end{equation} 
	where both $c_{(B)}$ and $d_{(B)}$ are nonlinear functions of
$B$, $P$, $Q$, and the radial derivatives of
$P$, and $Q$. We solved this equation in two different ways.
The first involves Newton's method: we define the vectors
$\mathcal{R}_j$ and $v_j$, with $0\geq j \geq N_r-1$ and
\begin{eqnarray}
	\mathcal{R}_j 
	\equiv
	\left(E^{(B)}\right)^{n+1}_{j+1/2}
	, \\
	v_j \equiv B^{n+1}_j
	,
\end{eqnarray}
	where $ (E^{(B)})^{n+1}_{j+1/2}$ is \eref{eq:EdGB_solved:rDer_B} with
the fields finite
differenced using the trapezoid stencil:
\begin{eqnarray}
\label{eq:trap_f_stencil}
	f
	& \to
	\frac{1}{2}\left(f^{n+1}_{j+1}+f^{n+1}_j\right)
	, \\
\label{eq:trap_rDer_f_stencil}
	\partial_rf
	& \to
	\frac{1}{\Delta r}\left(f^{n+1}_{j+1}-f^{n+1}_j\right)
	.
\end{eqnarray} 
\Eref{eq:EdGB_solved:rDer_B} is nonlinear in $B$, so we iteratively
solve for $B_j$ by solving for the linear correction $\delta v_j$
in
\begin{equation}
	\mathcal{J}_{ij}\delta v_j + \mathcal{R}_i
	= 0
	.
\end{equation}
	for $\delta v_j$, where 
\begin{equation}
	\mathcal{J}_{ij}
	\equiv
	\frac{\delta \mathcal{R}_i}{\delta v_j}
	.
\end{equation}
	As in
\ref{sec:hyperbolic_solvers_iter}
we inverted $\mathcal{J}_{ij}$ two different ways: one using a
banded matrix solver, and another iteratively using a Gauss-Seidel method.
The Newton iteration was then repeated until the 
residual 
$\mathcal{R}_j$ was below some tolerance well below truncation error.

	We also directly solved \Eref{eq:numerical_method_rDer_B} using a
second order Runge-Kutta method, by writing the equation as
$\partial_rB = - d/c$.

\subsection{\label{sec:ode_solvers_A}ODE integrator for $A$}
	The ODE for the $A$ field, (\ref{eq:EdGB_solved:rDer_A}) is of the form
\begin{equation}
	c_{(A)} \partial_rA + d_{(A)} = 0
	,
\end{equation}
	where $c_{(A)}$ and $d_{(A)}$
are functions of $B$, $P$, $Q$, and their radial
derivatives. We discretize the fields and their derivatives using
the trapezoidal rule as above (\ref{eq:trap_f_stencil},\ref{eq:trap_rDer_f_stencil}).
Since the ODE for $A$ is linear it is trivial to directly integrate it from the origin $j=0$
outward; specifically
we directly solve
for $A^{n+1}_{j+1}$ knowing $A^{n+1}_j$ and the other field values via
\begin{equation}
	A^{n+1}_{j+1}
	=
	A^{n+1}_j
	- \Delta r
	\frac{(d_{(A)})^{n+1}_{j+1/2}}{(c_{(A)})^{n+1}_{j+1/2}}
	,
\end{equation}
\subsection{\label{sec:partially_constrained_evo}Partially Constrained Evolution}
	In a partially constrained evolution, one (or more) variables
are typically solved for using an evolution instead of constraint
equation. Here, one can do that for $B$, with \eref{eq:EdGB_EOM:ThTh}
the corresponding second-order-in-time evolution equation for it. 
However, in Schwarzschild-like coordinates the 
momentum constraint \eref{eq:EdGB_EOM:TR} is effectively
a ``first integral'' for this equation, and instead then we
consider this as our evolution equation for $B$
(recall for our constrained
evolution we do not use the plain form of the momentum constraint,
but first eliminate the time derivative of $B$ using the other equations).
For initial data,
we solve for $B$ using 
\eref{eq:EdGB_solved:rDer_B} with
either an RK2 or a relaxation method. Once we begin evolving in time, we then
use \eref{eq:EdGB_solved:rDer_B} as an independent residual to monitor the
constraint.

	We solved a discretized  version of \eref{eq:EdGB_EOM:TR} for $B$ using an iterative Crank-Nicolson
method. On any given time step, we follow a similar procedure as 
above for the iterative constrained scheme, but now iterate over the evolution
equations for $Q^{n+1}$, $P^{n+1}$, and $B^{n+1}$ a fixed number of times,
then solve for $A^{n+1}$ using the constraint
equation, \eref{eq:EdGB_solved:rDer_A}. We repeat this process until the
residuals of the evolutions equations for $Q$, $P$, and $B$ are below
a tolerance set to be a few orders of magnitude below the truncation error.
Afterward we apply a Kreiss-Oliger filter on the variables
$Q^{n+1}$, $P^{n+1}$, and $B^{n+1}$, before advancing to the next time step.
	
\section*{References}

\bibliography{TEdGB_polarCoordinate_general}

\end{document}